\RequirePackage{fix-cm}
\documentclass[twocolumn,epjc3]{svjour3}
\RequirePackage{flushend}
\RequirePackage[colorlinks,citecolor=blue,urlcolor=blue,linkcolor=blue]{hyperref}
\RequirePackage{cite}
\RequirePackage{graphicx}
\RequirePackage{amssymb}
\RequirePackage{amsmath}
\begin{document}

\title{
Bound states of a fermion-dyon system
}
\author{A.Yu.~Loginov\thanksref{addr1,e1}}
\thankstext{e1}{e-mail: a.yu.loginov@tusur.ru}
\institute{Laboratory of Applied Mathematics and Theoretical Physics, Tomsk State
           University of Control Systems and Radioelectronics, 634050 Tomsk, Russia \label{addr1}}
\date{Received: date / Accepted: date}
\maketitle

\begin{abstract}
The bound states of fermions  in  the  external  field  of  an Abelian dyon are
studied here both analytically and numerically.
Their  existence  is  due  to  the  dyon's  electric  charge  resulting  from a
polarization of the fermionic vacuum.
The configuration of the dyon's field  is  not  invariant  under  $P$  or  $CP$
transformations.
The dependence of the energy levels of the fermion-dyon  system  on a parameter
of $CP$ violation is investigated.
The absence of $P$ invariance results in nonzero electric dipole moments of the
bound fermionic states.
These   depend    nontrivially    on   the    parameter   of   $CP$  violation.
The  bound  fermionic states  also  possess  nonzero  magnetic  dipole moments.
Unlike the electric dipole moments, the magnetic dipole moments are practically
independent of the parameter of $CP$ violation.
In addition, the magnitudes of the electric dipole moments significantly exceed
those of the magnetic dipole moments.

\end{abstract}

\section{Introduction} \label{sec:I}

The  interaction  of  fermions   with   the   Dirac  monopole \cite{Dirac_1931,
Dirac_1948} has  been  discussed  by  many authors \cite{Tamm_1931, fierz_1944,
 band_1946, harish_1948,      hurst_1968,      berr_1970,     wu_yang_npb_1976,
 wu_yang_prd_1977,      kzm1_prd_1977,     kzm2_prd_1977,   goldhaber_prd_1977,
 kzm3_prd_1977,  callias_prd_1977,   yamagishi_prd_1983,   osland_wu_npb_1984a,
 osland_wu_npb_1984b,          osland_wu_npb_1985a,        osland_wu_npb_1985b,
 osland_wu_npb_1985c,   osland_wu_npb_1985d,   bose_jpg_1986,   zhang_plb_1984,
 zhang_prd_1986, zhang_prd_1989, zhang_jmp_1990, zhang_prd_1990, zhang_jpa_1993,
 zhang_plb_2002, shnir_jpg_1988, shnir_ijmpa_1992, shnir_phscripta_1996}.
One reason for this is that the characteristic features of the fermion-monopole
system distinguish it from the hydrogen atom.
In  particular,  unlike a spherically symmetric electric field, the  monopole's
magnetic field is not invariant (it changes sign) under $P$ transformation.
As a result, bound  states  of the fermion-monopole  system may possess nonzero
electric dipole moments \cite{goldhaber_prd_1977,  kzm3_prd_1977}.
Another important  feature  is  the absence  of  the centrifugal barrier in the
state  with  the minimal angular momentum, which results in the fermions easily
reaching the location of the monopole.
Due to this, the Dirac Hamiltonian is  not self-adjoint on the subspace of wave
functions with the minimal angular momentum, which is unacceptable.
To solve this problem, an infinitesimally  small  ``extra''  magnetic moment is
added to the  Dirac fermion \cite{kzm1_prd_1977, kzm2_prd_1977, kzm3_prd_1977},
which is equivalent to a boundary condition at $r = 0$.

However, this boundary condition  is  not  the  most  general,  and  there is a
one-parameter     family      of      appropriate      boundary      conditions
\cite{goldhaber_prd_1977,  callias_prd_1977}, corresponding to the existence of
$\theta$ vacua.
Unlike  the  magnetic  monopole's  field,  these  boundary  conditions  are not
$CP$-invariant.
For massive fermions, this  leads  to  a violation  of  $CP$  invariance of the
fermion-monopole system.
It was  shown  in  \cite{yamagishi_prd_1983}  that  this  violation  leads to a
polarization  of  the fermionic vacuum in the vicinity of the  monopole,  as  a
result of   which   the   monopole   becomes  a  dyon whose electric charge  is
determined  by  the Witten formula \cite{witten_plb_1979}.
The long-distance asymptotics of the  dyon's  electric  potential is Coulombic,
but not its short-distance asymptotics.

The fermion-monopole     system      has     a      single      bound     state
\cite{goldhaber_prd_1977, callias_prd_1977}   in  a  specified  region  of  the
parameter $\theta$. 
This state possesses the minimal angular  momentum, and is, in general, tightly
bound.
The attractive Coulomb asymptotics of the dyon's  electric potential completely
changes this picture.
Like the hydrogen atom,  the  fermion-dyon  system  has  an  infinite number of
loosely bound states for  each  value  of  the  angular  momentum including the
minimal value.
As  with   a   purely   Coulomb   fermion-dyon   system   \cite{zhang_prd_1986,
zhang_prd_1989,zhang_jmp_1990,zhang_prd_1990},  the  energy  spectrum  of these
loosely bound states is hydrogen-like.
However,  in  contrast  to  \cite{zhang_prd_1986,zhang_prd_1989,zhang_jmp_1990,
zhang_prd_1990}, a twofold degeneracy of the energy  levels  is removed already
at the quantum mechanical level.

In general, the  bound  fermionic  states  possess  both  electric and magnetic
dipole moments.
The exception  is  the  bound  states   with   the  minimal  angular  momentum.
Their magnetic dipole moments vanish in any case, whereas their electric dipole
moments vanish only if the minimal angular momentum is zero.
The electric dipole moments  depend  nontrivially  on  the  parameter $\theta$,
whereas the magnetic dipole moments are practically independent of it.
The magnitudes of the electric dipole moments significantly exceed those of the
magnetic dipole moments.

This  paper  is  structured as follows.
In  Section~\ref{sec:II},  we   briefly  describe   some   properties   of  the
fermion-monopole system.
In Section~\ref{sec:III}, we describe the  polarization of the fermionic vacuum
in the  vicinity  of  the  monopole  and  the  characteristics  of  the induced
electric charge distribution.
In Section~\ref{sec:IV}, we study  the bound fermionic  states  arising  due to
the induced electric charge of the dyon.
We  consider  separately  the  bound  fermionic  states  with  the  minimal and
nonminimal angular momenta.
In Sections~\ref{sec:V}  and  \ref{sec:VI},  we study the electric and magnetic
dipole moments  of  the  bound fermionic states, respectively.
In the last section, we  briefly list the features of  the  fermion-dyon system
and summarise the results obtained in the present work.

Throughout this paper, the natural units $c = 1$ and $\hbar = 1$ are used.

\section{Some properties of the fermion-monopole system}         \label{sec:II}

The dynamics of a fermion in a fixed  Abelian  monopole field  is determined by
the Dirac equation
\begin{equation}
i \partial_{t} \psi  = H \psi,                                     \label{II:1}
\end{equation}
where the Hamiltonian
\begin{equation}
H = \boldsymbol{\alpha \cdot }\left(-i\nabla-e\mathbf{A}\right) + \beta M,
                                                                   \label{II:2}
\end{equation}
the matrices
\begin{equation}
\alpha^{i} = \gamma^{0}\gamma^{i},\; \beta = \gamma^{0},           \label{II:3}
\end{equation}
and we use the Dirac matrices in the standard representation
\begin{equation}
\gamma ^{0}=
\begin{pmatrix}
1 & 0 \\
0 & -1
\end{pmatrix}
,\;\gamma ^{i}=
\begin{pmatrix}
0 & \sigma ^{i} \\
-\sigma ^{i} & 0
\end{pmatrix}.                                                     \label{II:4}
\end{equation}
In Eq.~(\ref{II:1}), we consider the monopole vector potential $\mathbf{A}$ not
as an ordinary  vector  field,  but  as  a connection  on  a nontrivial  $U(1)$
bundle over $R^{3}/\left\{0\right\} \sim S^{2}$.
Accordingly, the wave function $\psi$ should  be considered as a section rather
than an ordinary function.
This approach, proposed in \cite{wu_yang_1975}, allows one to avoid an unwanted
string singulary when describing the monopole.

To describe the monopole's vector potential we must consider the space outside
of the monopole as the  union  of  two overlapping regions $R_{a}$ and $R_{b}$,
\begin{subequations}                                               \label{II:5}
\begin{eqnarray}
R_{a}\! &:&\;r>0,\;0\leq \vartheta \leq \frac{\pi }{2}+\delta ,\;0\leq
\varphi <2\pi ,                                                   \label{II:5a}
 \\
R_{b}\! &:&\;r>0,\;\frac{\pi }{2}-\delta \leq \vartheta \leq \pi ,\;0\leq
\varphi <2\pi ,                                                   \label{II:5b}
\end{eqnarray}
\end{subequations}
where the angle $\delta$ is in the range $\left(0, \pi/2\right)$.
Then, in the regions $R_{a}$  and  $R_{b}$,  the  monopole vector  potential is
chosen to be
\begin{equation}
\mathbf{A}^{\left( a\right) }=-\frac{g}{r}\frac{\mathbf{r}\times \mathbf{n}}
{r+\mathbf{r \cdot n}}                                             \label{II:6}
\end{equation}
and
\begin{equation}
\mathbf{A}^{\left( b\right) }=\frac{g}{r}\frac{\mathbf{r}\times \mathbf{n}}
{r-\mathbf{r \cdot n}},                                            \label{II:7}
\end{equation}
respectively, where $g$ is the monopole's  magnetic  charge and the unit vector
$\mathbf{n} = (0, 0, 1)$.
Note that both  $\mathbf{A}^{\left(a\right)}$ and $\mathbf{A}^{\left(b\right)}$
are  nonsingular  in  their   domains   of   definition  $R_{a}$  and  $R_{b}$.
In the region of overlap $R_{a}\cap R_{b}$, the two potentials are related by a
gauge transformation
\begin{equation}
A_{\mu }^{\left( a\right) }=A_{\mu }^{\left( b\right) }+\frac{i}{e}%
S_{ab}\partial _{\mu }S_{ab}^{-1},                                 \label{II:8}
\end{equation}
where the transition function
\begin{equation}
S_{ab} = e^{2 i e g\varphi}.                                       \label{II:9}
\end{equation}

Similar to the connection $\mathbf{A}$, the wave section $\psi$ is given by two
wave functions $\psi_{a}$  and  $\psi_{b}$  defined  in the regions $R_{a}$ and
$R_{b}$, respectively.
In the region $R_{a}$  ($R_{b}$),  the  wave  function  $\psi_{a}$ ($\psi_{b}$)
satisfies the Dirac equation (\ref{II:1}) with  the vector potential $\mathbf{A
}^{\left(a\right)}$ ($\mathbf{A}^{\left(b\right)}$).
In the region $R_{a} \cap R_{b}$, the wave  functions $\psi_{a}$ and $\psi_{b}$
are connected by the relation
\begin{equation}
\psi_{a} = S_{ab}\psi_{b}.                                        \label{II:10}
\end{equation}
The necessity of the single-valuedness  of  the transition function $S_{ab}$ in
the region  $R_{a}  \cap  R_{b}$  leads  to  the  Dirac  quantization condition
\cite{Dirac_1931}
\begin{equation}
q \equiv e g = \frac{n}{2},                                       \label{II:11}
\end{equation}
where $n$ is an integer.

The Hamiltonian  (\ref{II:2})  commutes  with  the angular momentum operator
\begin{equation}
\mathbf{J}=\mathbf{r\times }\left(-i\nabla-e\mathbf{A}\right)+\mathbf{S}
-q\frac{\mathbf{r}}{r},                                           \label{II:12}
\end{equation}
where the spin operator
\begin{equation}
\mathbf{S}=\frac{1}{2}%
\begin{pmatrix}
\boldsymbol{\sigma } & 0 \\
0 & \boldsymbol{\sigma }%
\end{pmatrix}.                                                    \label{II:13}
\end{equation}
A characteristic feature  of  operator  (\ref{II:12})  is  the  presence of the
radial term $q \mathbf{r}/r$, which  leads to a number of unusual properties of
the fermion-monopole system.
In particular, it follows  from  Eq.~(\ref{II:12})  and  the  addition rule for
angular momenta that  the  minimum  value  of  the  fermion angular momentum is
$j = \left\vert q\right\vert - 1/2$.
Hence, for the fundamental Dirac monopole ($n=1,\,q=1/2$), the minimum value of
the fermion angular momentum is zero.
Furthermore, a fermion-monopole system  with an odd (even) $n$ can possess only
integer (half-integer) angular momenta.

\subsection{\label{subsec:IIA} Properties of  the fermion-monopole system under
the discrete transformations of QFT}

Now we shall investigate the properties  of  the  fermion-monopole system under
the discrete transformations of QFT.
First, we note that the two components  of  the connection $\mathbf{A}$ satisfy
the relation
\begin{equation}
\mathbf{A}^{\left(a, b\right)}\left(\mathbf{x}\right) =
\mathbf{A}^{\left(b,a\right) }\left( -\mathbf{x}\right).          \label{II:14}
\end{equation}
Then, it follows  from  Eqs.~(\ref{II:6}), (\ref{II:7}), and (\ref{II:14}) that
under  $P$   transformation,  $\mathbf{A}^{\left(a,  b\right) }\left(\mathbf{x}
\right)\!\rightarrow\!\mathbf{A}^{P\left( a,b\right) }\left(\mathbf{x}\right)$,
where
\begin{equation}
\mathbf{A}^{P\left( a,b\right) }\left( \mathbf{x}\right)\!=\!-\mathbf{A
}^{\left(b,a\right)}\left(-\mathbf{x}\right)\!=\!-\mathbf{A}^{\left(a,b
\right)}\left( \mathbf{x}\right).                                 \label{II:15}
\end{equation}
We see  that  $P$  transformation   is  equivalent  to   the  reversal  of sign
$g \rightarrow -g$ of the magnetic charge of the monopole.
Using Eqs.~(\ref{II:1}), (\ref{II:2}), (\ref{II:15}), and the properties of the
Dirac matrices, we obtain the transformation law for the fermion wave functions
$\psi_{\left(a,b\right)}$ under $P$  transformation
\begin{equation}
\psi_{\left(a,b\right)}^{P}\left(t,\mathbf{x}\right) = \eta_{P}\gamma
^{0}\psi_{\left(b,a\right)}\left(t,-\mathbf{x}\right),            \label{II:16}
\end{equation}
where $\eta_{P}$ is a phase factor.
The $P$ transformation is not a symmetry  of the fermion-monopole system, which
distinguishes it from the hydrogen atom.
Rather, it transforms the  fermion (antifermion) wave function  in the external
field of the monopole of charge $g$ to the fermion (antifermion)  wave function
in the external field of the monopole of charge $-g$.

Under  $C$  conjugation,  the  components  of  the  connection $\mathbf{A}$ are
transformed just as under $P$ transformation
\begin{equation}
\mathbf{A}^{\left( a,b\right) }\left( \mathbf{x}\right) \rightarrow \mathbf{A
}^{C\left( a,b\right) }\left( \mathbf{x}\right) =-\mathbf{A}^{\left(
a,b\right) }\left( \mathbf{x}\right),                             \label{II:17}
\end{equation}
whereas the fermion wave functions are transformed as
\begin{equation}
\psi_{\left(a,b\right)}^{C}\left( t,\mathbf{x}\right)=\eta _{C}\gamma
^{2}\psi _{\left( a,b\right)}^{\ast }\left( t,\mathbf{x}\right),  \label{II:18}
\end{equation}
where $\eta_{C}$ is a phase factor.
We see that  like $P$ transformation,  $C$ conjugation is not a symmetry of the
fermion-monopole system.
Instead, it transforms the  fermion (antifermion) wave function in the external
field of  the  monopole  of   charge  $g$  to  the  antifermion  (fermion) wave
function in the external field  of the monopole of charge $-g$.

Combining  Eqs.~(\ref{II:15})   and   (\ref{II:17}),  we   conclude  that  $CP$
inversion leaves the monopole's vector potential unchanged.
At the same time, Eqs.~(\ref{II:16}) and (\ref{II:18}) tell us that under $CP$,
the wave function $\psi_{\left(a, b\right)}\left(t,\mathbf{x}\right)\rightarrow
\psi_{\left(a, b\right)}^{CP}  \left(t, \mathbf{x}\right)$,
where
\begin{equation}
\psi _{\left( a,b\right) }^{CP}\left( t,\mathbf{x}\right) =\eta _{CP}\alpha
_{2}\psi_{\left(b,a\right) }^{\ast }\left( t,-\mathbf{x}\right),  \label{II:19}
\end{equation}
the $CP$ phase $\eta_{CP}=-\eta_{C}\eta_{P}^{\ast}$, and the matrix $\alpha_{2}
=\gamma^{0}\gamma^{2}$.
Since  $CP$  inversion  does  not change  the vector potential of the monopole,
it could be a symmetry of the fermion-monopole  system transforming its fermion
(antifermion) states to antifermion (fermion) ones.
However, we shall see later that in the general case, $CP$ is not a symmetry of
the fermion-monopole system.

\subsection{\label{subsec:IIB}      The fermion-monopole system in a state with
the lowest angular momentum $j = \left\vert q \right\vert - 1/2$}

It was  shown  in  \cite{kzm1_prd_1977}  that  the  angular  momentum  operator
(\ref{II:12}) has three types of eigensections.
We focus on the eigensection  of  the third  type, since it describe the lowest
partial wave with $j = \left\vert q \right\vert - 1/2$.
The fermion wave section is written as
\begin{equation}
\psi _{j m}^{\left( 3\right) }\left( t,\mathbf{x}\right) =\frac{1}{r}
\chi \left( r\right) \otimes \eta _{j m}\left( \vartheta ,\varphi
\right)e^{-iEt},                                                  \label{II:20}
\end{equation}
where the radial part is
\begin{equation}
\chi \left( r\right) =
\begin{bmatrix}
f\left( r\right)  \\
g\left( r\right)
\end{bmatrix},                                                    \label{II:21}
\end{equation}
the eigensection of the third type is
\begin{equation}
\eta _{j m}=
\begin{bmatrix}
-\left( \dfrac{j-m+1}{2 j + 2}
\right) ^{1/2}Y_{q,\left\vert q\right\vert, m-1/2} \\
\left( \dfrac{j+m+1}{2 j + 2}
\right) ^{1/2}Y_{q,\left\vert q\right\vert ,m+1/2}
\end{bmatrix},                                                    \label{II:22}
\end{equation}
the angular momentum $j = \left\vert q \right\vert-1/2$, and the $z$ projection
$m$  of  the  angular  momentum is in the range $\left[-\left\vert q\right\vert
+1/2,\left\vert q\right\vert -1/2\right]$.
The monopole spherical  harmonics $Y_{q,l,\mu}\left( \vartheta ,\varphi\right)$
entering into Eq.~(\ref{II:22})  were  defined  in \cite{wu_yang_npb_1976}, and
their properties were derived in \cite{wu_yang_prd_1977}.
In Eq.~(\ref{II:20}), the nontrivial topology of the  monopole $U(1)$ bundle is
completely reflected  in  the  eigensection $\eta_{j m}\left(\vartheta, \varphi
\right)$, whereas  $\chi\left( r \right)$  is  an  ordinary  two-component wave
function defined on the interval $\left[0,\infty\right)$.

Substituting Eq.~(\ref{II:20}) into Eq.~(\ref{II:1}) and  using  the properties
\cite{kzm1_prd_1977} of  the  eigensection  $\eta_{j m}\left(\vartheta, \varphi
\right)$, we reduce  the  Dirac  equation  to  a  system  of  two  differential
equations for the radial functions 
\begin{equation}
H\chi \left(r\right) = E\chi \left(r\right),                      \label{II:23}
\end{equation}
where the reduced Hamiltonian is
\begin{equation}
H=-i\frac{q}{\left\vert q\right\vert }\tilde{\gamma}_{5}\frac{d}{dr}
+M \tilde{\beta}                                                  \label{II:24}
\end{equation}
with the matrices
\begin{equation}
\tilde{\gamma}_{5} =
\begin{bmatrix}
0 & 1 \\
1 & 0
\end{bmatrix}
\;\;\text{and}\;\;
\tilde{\beta} =
\begin{bmatrix}
1 & 0 \\
0 & -1
\end{bmatrix}.                                                    \label{II:25}
\end{equation}
Using Eqs.~(\ref{II:15}) -- (\ref{II:22}), it can be shown that the  effects of
$P$ and $CP$ transformations on  the  two-component radial wave function $\chi$
are
\begin{align}
& \chi \overset{P}{\rightarrow}\chi ^{P} = \tilde{\beta}\chi,\;
q \overset{P}{\rightarrow}-q,\;
E \overset{P}{\rightarrow} E                                     \label{II:26a}
 \\
& \text{and}                                                       \nonumber
 \\
& \chi\overset{CP}{\longrightarrow}\chi^{CP}=i \tilde{\gamma}_{5}\chi^{\ast},\;
q \overset{CP}{\longrightarrow} q, \;
E \overset{CP}{\longrightarrow}-E,                               \label{II:26b}
\end{align}
respectively.
Eq.~(\ref{II:26a}) tells us that we  may  restrict ourselves to the case $q>0$,
since the solutions for $q < 0$  can  be obtained by the parity transformation.

The reduced Hamiltonian (\ref{II:24}) corresponds to  the state with the lowest
possible angular momentum $j = \left\vert q \right\vert - 1/2$.
Its characteristic feature  is  the  absence  of a centrifugal barrier at small
$r$,  which  is  a  consequence  of   the   presence   of  the  additional term
$-q \mathbf{r}/r$ in Eq.~(\ref{II:12}).
The two-component radial wave functions $\chi\left(r\right)$ are defined on the
interval $\left[0,\infty\right)$, and their inner product is defined by
\begin{equation}
\left( \chi _{1},\chi _{2}\right) =\int\nolimits_{0}^{\infty }\chi
_{1}^{\dagger }\left( r\right) \chi _{2}\left( r\right) dr.       \label{II:27}
\end{equation}
To correspond to a physical observable, the  Hamiltonian  (\ref{II:24}) must be
Hermitian with respect to this inner product, i.e,
\begin{eqnarray}
&&\left(\chi_{1},H\chi_{2}\right) - \left(H\chi_{1},\chi_{2}\right) =
i\chi_{1}^{\dag}\left(0\right)\tilde{\gamma}_{5}
\chi_{2}\left( 0\right) =                                           \nonumber
  \\
&&i\left[ f_{1}^{\ast }\left( 0\right) g_{2}\left( 0\right) +g_{1}^{\ast
}\left( 0\right) f_{2}\left( 0\right) \right] = 0,                \label{II:28}
\end{eqnarray}
where we  have  assumed  that  $\chi_{1}$  and  $\chi_{2}$  vanish sufficiently
rapidly at infinity.

The simple structure of  the  Hamiltonian (\ref{II:24}) allows us to obtain the
general solution to Eq.~(\ref{II:23}) in an analytical form
\begin{equation}
\chi _{E}\left( r\right) =c_{1}\chi _{E}^{\left( 1\right) }\left( r\right)
+c_{2}\chi _{E}^{\left( 2\right) }\left( r\right),                \label{II:29}
\end{equation}
where
\begin{align}
&\chi _{E}^{\left( 1\right) }\left( r\right)  =
\begin{bmatrix}
\dfrac{ik}{E-M}\sin \left(k r\right)  \\
\cos \left( kr\right)
\end{bmatrix},                                                    \label{II:30}
\\
&\chi _{E}^{\left( 2\right) }\left( r\right)  =
\begin{bmatrix}
\cos \left( kr\right)  \\
\dfrac{ik}{E+M}\sin \left(k r\right)
\end{bmatrix},                                                    \label{II:31}
\\
&E = \pm \left(k^{2} + M^{2}\right)^{1/2},                        \label{II:32}
\end{align}
and it is assumed that $k > 0$.
It follows from Eqs.~(\ref{II:29}) -- (\ref{II:31})  that  $\chi_{E}^{\text{T}}
\left(0\right)=\left(c_{1},c_{2}\right)$, and it does not vanish except for the
trivial case $c_{1} = c_{2} = 0$.
Hence, in general, the  Hermiticity  condition  (\ref{II:28}) is not satisfied,
and the Hamiltonian is  not  a  self-adjoint  operator  on  the states with the
lowest angular momentum $j = \left\vert q \right\vert - 1/2$.
Furthermore, it can be shown that the radial  component  of the fermion current
$j^{\mu}  =  \bar{\psi}^{(3)}\gamma ^{\mu } \psi^{(3)}$  is  $j_{r} = \text{Re}
\left[c_{1}c_{2}^{\ast }\right] /\left( 2\pi r^{2}\right)$.
It follows that the fermion flux  through  a  spherical surface surrounding the
monopole is $\mathrm{\Phi} = 2 \text{Re} \left[c_{1} c_{2}^{\ast }\right]$, and
is not equal to zero in the general case.
This is equivalent to the presence of a source of fermions (or antifermions) at
the origin and contradicts the unitarity condition.

Nevertheless,  the  unitarity  problem  will  be  solved  and  the  Hermiticity
condition  (\ref{II:28}) will be satisfied provided that for any two  solutions
$\chi_{1}^{\text{T}}(r) =  (f_{1}(r),g_{1}(r))$  and  $\chi_{2}^{\text{T}}(r) =
(f_{2}(r), g_{2}(r))$  of  Eq.~(\ref{II:23})  (including the coincident case of
$\chi_{1}(r)=\chi_{2}(r)$), the sesquilinear  combination  $f_{1}^{\ast}\left(0
\right)g_{2}\left( 0 \right)+g_{1}^{\ast}\left( 0 \right)f_{2}\left( 0 \right)$
vanishes, which is equivalent to the relation $\left[f_{1} \left(0\right)/g_{1}
\left(0\right)\right]^{\ast} = -f_{2}\left(0\right) /g_{2}\left(0\right)$.
It follows that  any  solution $\chi^{\text{T}}\left(r\right) = \left( f\left(r
\right),  g\left( r \right) \right)$   of    Eq.~(\ref{II:23})   must   satisfy
\cite{goldhaber_prd_1977, callias_prd_1977}
\begin{equation}
\frac{f\left( 0\right) }{g\left( 0\right) }=i\tan \left[ \frac{\theta }{2}+
\frac{\pi }{4}\right],                                            \label{II:33}
\end{equation}
where the parametric angle $\theta \in \left( -\pi ,\pi \right)$.
Thus, the formal Hamiltonian  (\ref{II:24})  admits  a  one-parameter family of
self-adjoint  extensions  \cite{goldhaber_prd_1977, callias_prd_1977}  provided
that its eigenfunctions satisfy Eq.~(\ref{II:33}).

It  can  be  easily  checked  that  under  $CP$  inversion (\ref{II:26b}),  the
parameter $\theta$ changes sign
\begin{equation}
\theta \overset{CP}{\longrightarrow }-\theta.                    \label{II:33a}
\end{equation}
It follows  that  in  the  general  case, $CP$   is   not   a  symmetry  of the
fermion-monopole system \cite{yamagishi_prd_1983}.
The only  exceptions  are  for  the  values  of  the  parameter  $\theta = 0,\,
\text{and}\, \pm \pi$.
In the latter case, the  values  of  $\theta = \pm  \pi$ correspond to the same
state, since $\tan \left[-\pi /2+\pi /4\right]=\tan \left[\pi /2 + \pi/4\right]
= -1$ and condition (\ref{II:33}) remains unchanged.

Note that there are no problems related  to  the  Hermiticity and unitarity for
states of the fermion-monopole system with the angular momentum $j > \left\vert
q \right\vert - 1/2$.
This is because there is a centrifugal barrier in  this case, and therefore the
radial wave functions vanish at the origin.

\subsection{\label{subsec:IIC} Eigenfunctions of the Hamiltonian in the case of
massive fermions}

In the  case  of  massive  fermions,  the  reduced   Hamiltonian  (\ref{II:24})
possesses   the   following   eigenfunctions   of    the   continuous  spectrum
\cite{yamagishi_prd_1983}:
\begin{eqnarray}
u_{k \theta }\left( r\right)  &=&\frac{2^{1/2}k}{\left[ E\left( E-M\sin
\left( \theta \right) \right) \right] ^{1/2}}                       \nonumber
 \\
&&\times \left[ \cos \left( \frac{\theta }{2}+\frac{\pi }{4}\right) \chi
_{E}^{\left( 1\right) }\left( r\right) \right.                    \nonumber
 \\
&&+ \left. i\sin \left( \frac{\theta }{2}+\frac{\pi }{4}\right) \chi
_{E}^{\left( 2\right) }\left( r\right) \right]                    \label{II:34}
\end{eqnarray}
for $E=\left( k^{2}+M^{2}\right) ^{1/2}$, and
\begin{eqnarray}
v_{k \theta }\left( r\right)  &=&\frac{2^{1/2}k}{\left[ \left\vert
E\right\vert \left( \left\vert E\right\vert +M\sin \left( \theta \right)
\right) \right] ^{1/2}}                                             \nonumber
 \\
&&\times \left[ \cos \left( \frac{\theta }{2}+\frac{\pi }{4}\right) \chi
_{E}^{\left( 1\right) }\left( r\right) \right.                      \nonumber
 \\
&&+ \left. i\sin \left( \frac{\theta }{2}+\frac{\pi }{4}\right) \chi
_{E}^{\left( 2\right) }\left( r\right) \right]                    \label{II:35}
\end{eqnarray}
for $E=-\left( k^{2}+M^{2}\right) ^{1/2}$.
In addition,  there   is   also   a   bound   state   \cite{goldhaber_prd_1977,
 callias_prd_1977} of the discrete spectrum  with the energy $E = M \sin \left(
\theta \right)$, provided that  $\cos \left( \theta \right) < 0 \Leftrightarrow
\theta  \in  \left[-\pi, -\pi /2 \right) \cup \left(\pi /2,\pi \right]$.
The radial wave function of this state is
\begin{equation}
B_{\theta }\left( r\right) =
\begin{bmatrix}
i\sin \left( \dfrac{\theta }{2}+\dfrac{\pi }{4}\right) \\
\cos \left( \dfrac{\theta }{2}+\dfrac{\pi }{4}\right)
\end{bmatrix}%
\sqrt{2\kappa }e^{-\kappa r},                                     \label{II:36}
\end{equation}
where the parameter $\kappa=M\left\vert \cos \left( \theta\right) \right\vert$.

The eigenfunctions (\ref{II:34}) -- (\ref{II:36}) satisfy
\begin{subequations}                                              \label{II:37}
\begin{eqnarray}
\left( u_{k \theta },u_{k^{\prime }\theta }\right)  &=&2\pi \delta \left(
k-k^{\prime }\right),                                            \label{II:37a}
  \\
\left( v_{k \theta },v_{k^{\prime }\theta }\right)  &=&2\pi \delta \left(
k-k^{\prime }\right),                                            \label{II:37b}
  \\
\left( B_{\theta },B_{\theta }\right)  &=&1,                     \label{II:37c}
\end{eqnarray}
\end{subequations}
while all other inner products vanish.
We  see  that  eigenfunctions  (\ref{II:34}) -- (\ref{II:36})  form  a complete
orthonormal system on the half-line $r \ge 0$.
Finally, it is easily shown that under $CP$ inversion
\begin{subequations}                                              \label{II:38}
\begin{eqnarray}
u_{k\theta }\overset{CP}{\longrightarrow }u_{k\theta }^{CP}
&=&i\tilde{\gamma}_{5}u_{k \theta }^{\ast}=v_{k\,-\theta },      \label{II:38a}
 \\
v_{k \theta }\overset{CP}{\longrightarrow }v_{k \theta }^{CP}
&=&i\tilde{\gamma}_{5}v_{k \theta }^{\ast }=u_{k\,-\theta },     \label{II:38b}
 \\
B_{\theta }\overset{CP}{\longrightarrow }B_{\theta }^{CP} &=&i\tilde{\gamma}
_{5}B_{\,\theta }^{\ast }=B_{\,-\theta },                        \label{II:38c}
\end{eqnarray}
\end{subequations}
in accordance with Eq.~(\ref{II:33a}).

\section{Polarization  of  the   fermionic   vacuum   in  the  vicinity  of the
magnetic monopole}                                              \label{sec:III}

It follows from Eqs.~(\ref{II:38a}) -- (\ref{II:38c}) that the fermion-monopole
system  is  not  $CP$-invariant,  since  the parameters $\pm \theta$ correspond
to two different fermion-monopole systems.
It was  shown   in   \cite{yamagishi_prd_1983}  that  this  violation  of  $CP$
invariance results in a polarization of the fermionic vacuum in the vicinity of
the monopole, with the result that the monopole becomes a dyon.
The density of the induced electric charge is
\begin{eqnarray}
\rho \left( r,\theta \right)  &=&-\frac{qeM\sin \left( \theta \right)}
{2\pi^{2}r^{2}}                                                     \nonumber
  \\
&&\times
\int\nolimits_{M}^{\infty }\frac{\kappa \exp \left( -2\kappa r\right) }{
\left( \kappa ^{2}-M^{2}\right) ^{1/2}\left( \kappa +M\cos \left( \theta
\right) \right) }d\kappa.                                         \label{III:1}
\end{eqnarray}
Although the integral in Eq.~(\ref{III:1}) cannot  be  calculated  analytically
in the general case, the induced electric charge $Q = 4 \pi \int\nolimits_{0}^{
\infty}\rho\left( r,\theta\right) r^{2}dr$ can be obtained   in   an analytical
form
\cite{yamagishi_prd_1983}
\begin{equation}
Q=-\frac{q e\theta}{\pi}.                                         \label{III:2}
\end{equation}
For a monopole with minimal magnetic charge ($n=1,\,q =1/2$), Eq.~(\ref{III:2})
is precisely the Witten formula \cite{witten_plb_1979, wilczek_prl_1982}.
We conclude that for $M > 0$, the  Abelian magnetic monopole becomes a dyon due
to the polarization of the fermionic vacuum.

Using Eq.~(\ref{III:1}),   one   can   derive   several  analytical expressions
related to the spatial distribution of the dyon's electric charge.
In particular,   the   mean  radius of the  charge  distribution  is $\langle r
\rangle_{\text{el}}=4 \pi Q^{-1}\int\nolimits_{0}^{\infty}r \rho\left(r, \theta
\right) dr$  and  the  mean  radius  squared   of  the  charge  distribution is
$\langle r^{2}\rangle_{\text{el}} = 4 \pi Q^{-1}\int\nolimits_{0}^{\infty}r^{2}
\rho\left(r,\theta\right)dr$.
Both can be expressed in terms of elementary functions:
\begin{equation}
\langle r \rangle_{\text{el}}=\frac{1}{2M\cos \left( \theta \right) }
\left( \frac{\pi \sin \left( \theta \right) }{2\theta}-1\right)  \label{III:2b}
\end{equation}
and
\begin{equation}
\langle r^{2}\rangle_{\text{el}}=\frac{1}{4M^{2}\cos ^{2}\left(\theta\right)}
\left( 2-\frac{\pi \sin \left( \theta \right) }{\theta }+\frac{\sin \left(
2\theta \right) }{\theta }\right).                               \label{III:2c}
\end{equation}
Using Eqs.~(\ref{III:2b}) and (\ref{III:2c}), we then obtain the expression for
the dispersion $D_{\text{el}}  =  \langle r^{2} \rangle_{\text{el}} - \langle r
\rangle_{\text{el}}^{2}$  of the charge distribution
\begin{equation}
D_{\text{el}}=\frac{1}{4M^{2}\cos^{2}\left(\theta \right)}\left(1-\frac{\pi^{2}
\sin^{2}\left( \theta \right) }{4\theta ^{2}}+\frac{\sin \left( 2\theta \right)
}{\theta }\right).                                               \label{III:2d}
\end{equation}

The electric potential corresponding to the charge density (\ref{III:1}) can be
written as
\begin{eqnarray}
A_{0}\left( r\right)  &=&-\frac{qe\theta }{\pi r}+\frac{qeM\sin \left(
\theta \right) }{\pi r}                                             \nonumber
  \\
&&\times
\int\nolimits_{M}^{\infty }\frac{\left( e^{-2\kappa r}+2\kappa r\text{Ei}
\left( -2\kappa r\right) \right) }{\left( \kappa ^{2}-M^{2}\right)
^{1/2}\left( \kappa +M\cos \left(\theta \right)\right)}d\kappa,  \label{III:2a}
\end{eqnarray}
where $\text{Ei}(x)$ is the exponential integral function \cite{DLMF}.
The definite integrals in Eqs.~(\ref{III:1}) and (\ref{III:2a}) cannot be found
in an analytical form in the general case.
We can,  however,   obtain  the   asymptotics  of  the  electric charge density
(\ref{III:1})  and   electric  potential (\ref{III:2a}) using standard methods:
\begin{eqnarray}
\rho \left( r, \theta \right)  &\sim &\frac{qeM\sin \left(\theta \right)}{
2\pi ^{2}r^{2}}\ln \left( Mr\right),                             \label{III:3a}
  \\
A_{0}\left(r,\theta\right)&\sim &-\frac{qeM\sin\left(\theta \right) }{\pi}
\left( 1-\ln \left(M r\right) \right)^{2}                       \label{III:3b}
\end{eqnarray}
for $r \rightarrow 0$, and
\begin{eqnarray}
\rho \left( r,\theta \right)  &\sim &-\frac{qeM\tan \left( \theta /2\right)
}{2\pi ^{2}r^{2}}\left( \frac{\pi }{4Mr}\right) ^{1/2}e^{-2Mr},  \label{III:4a}
 \\
A_{0}\left(r,\theta\right)  &\sim &-\frac{qe\theta }{\pi r}+\frac{ q e \tan
\left( \theta /2\right) }{\pi r}\frac{\sqrt{\pi }}{4}\frac{e^{-2Mr}}{
\left( Mr\right) ^{3/2}}                                         \label{III:4b}
\end{eqnarray}
for $r \rightarrow \infty$.
The Eqs.~(\ref{III:3a}) -- (\ref{III:4b}) become inapplicable for $\theta = \pm
\pi$.
In this case, we can obtain  an analytical form of $\rho\left(r,\pm \pi\right)$
after an integration by parts
\begin{equation}
\rho \left(r,\pm \pi \right) =\mp \frac{qeM}{2\pi r^{2}}e^{-2Mr}. \label{III:5}
\end{equation}
The electric potential $A_{0}(r,\pm \pi)$ can also be obtained in
an analytical form
\begin{equation}
A_{0}\left(r,\pm\pi\right) = \mp \frac{qe}{r}
\left[1\!-\!e^{-2Mr}\!\!+\!2Mr\mathrm{\Gamma}
\left(0, 2Mr\right)\right]\!,                                     \label{III:6}
\end{equation}
where $\mathrm{\Gamma} \left(0, 2 M r\right)$  is the incomplete gamma function
\cite{DLMF}.

Eqs.~(\ref{III:3a}), (\ref{III:4a}), and (\ref{III:5})  tell us that the radial
density $4\pi r^{2}\rho(r)$ has  an  integrable  singularity  at the origin and
tends to zero exponentially as $r \rightarrow \infty$.
Further, the characteristic size (\ref{III:2b})  of  the charge distribution is
on the order of the inverse fermion mass $M^{-1}$.
This smeared charge  distribution  results  in  the potential $A_{0}(r)$ having
only a weak logarithmic singularity at the origin.
In contrast, the point-like  ($\propto \delta(\mathbf{x})$) distribution of the
electric  charge  leads  to  a   pole  singularity  ($\propto r^{-1}$)  of  the
potential $A_{0}(r)$.
As a result, the electrostatic energy  of  the  charge distribution surrounding
the magnetic monopole is finite and can be written as
\begin{equation}
E_{\text{el}} = q^{2} \alpha M \tilde{E}_{\text{el}}\left(\theta\right),
                                                                 \label{III:6a}
\end{equation}
where $\alpha=e^{2}$ is the fine structure constant and  $\tilde{E}_{\text{el}}
\left( \theta\right)$ is an even dimensionless function.
For example, the  electrostatic  energy  corresponding to the parameter $\theta
= \pm \pi$ is
\begin{equation}
E_{\text{el}}(\pm\pi)=q^{2}\alpha M \ln\left(4\right).            \label{III:7}
\end{equation}

\begin{figure}[tbp]
\includegraphics[width=0.5\textwidth]{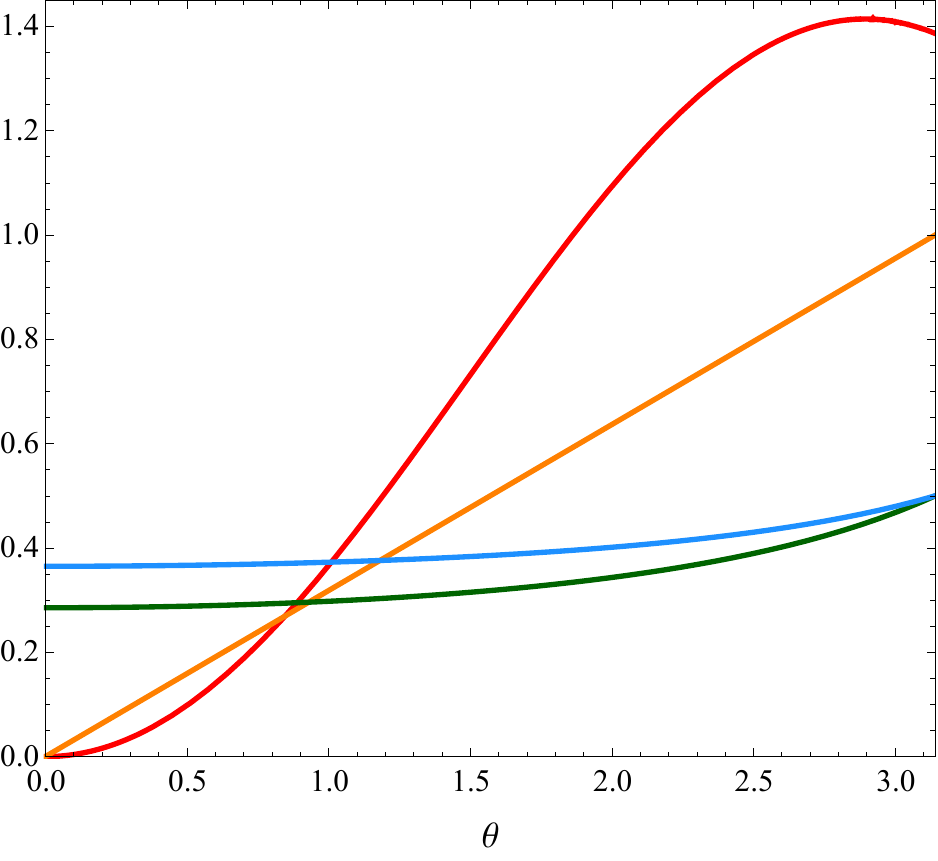}
\caption{\label{fig1}       Dependence   of   the   dimensionless   quantities
$\tilde{E}_{\text{el}}$    (red   curve),    $|\tilde{Q}|$    (orange   curve),
$\langle\tilde{r} \rangle_{\text{el}}$  (green  curve),  and  $\tilde{\sigma}_{
\text{el}}$ (blue curve) on the parameter $\theta$}
\end{figure}

Similar   to   $\tilde{E}_{\text{el}}\left( \theta \right)$,   we   define  the
dimensionless mean radius  $\langle \tilde{r} \rangle_{\text{el}} = M \langle r
\rangle_{\text{el}}$ and  the  scaled  electric charge $\tilde{Q} = Q/\left(q e
\right) = -\theta /\pi$.
We also define the dimensionless standard deviation $\tilde{\sigma}_{\text{el}}
=M \sigma_{\text{el}}=M D_{\text{el}}^{1/2}$, which characterizes the deviation
from the mean radius.
Figure~\ref{fig1}  shows  the  dependence  of these dimensionless quantities on
the parameter $\theta$.
The most interesting feature is that $\tilde{E}_{\text{el}}\left(\theta\right)$
is not a monotonically increasing function of $\theta$.
Instead, $\tilde{E}_{\text{el}}\left( \theta \right)$ reaches  a global maximum
approximately equal to $1.41$ at $\theta \approx 2.9$.
Hence, the electrostatic energy (\ref{III:6a}) is not a monotonically increasing
function  of the magnitude  of  the electric charge $\left\vert Q \right\vert$,
since the latter is related to $\theta$ by  the linear  relation  $\left\vert Q
\right\vert=\text{sgn}\left(\theta \right)\left\vert q\right\vert e\theta/\pi$.
This is because the effective size of charge  distribution (\ref{III:1}), which
is characterized by the parameters  $\langle \tilde{r} \rangle_{\text{el}}$ and
$\tilde{\sigma}_{\text{el}}$,  increases   monotonically  with  an  increase in
$\theta$.
As a result, for $\theta \gtrsim 2.9$,  the  effect  of  increasing the spatial
size of  the  charge  distribution  overpowers  the  effect  of  increasing the
magnitude of  the  electric  charge,  and  the  electrostatic  energy begins to
decrease.

It should be noted  that  the values of $\theta  =  \pm  \pi$  require  special
consideration, since  the  energy  $E = M \sin(\theta)$  of the bound fermionic
state vanishes in this case.
Here we deal with the  presence  of  the  fermionic  zero  modes  $B_{\pm \pi}$
in the external  field   of   a  topologically  nontrivial  field configuration
(the  monopole bundle).
However, it follows from Eqs.~(\ref{II:36}) and (\ref{II:38c}) that
\begin{equation}
B_{\pi }\overset{CP}{\longrightarrow }B_{-\pi } = -B_{\pi}.       \label{III:8}
\end{equation}
We see that the zero modes  $B_{\pi }$ and $B_{-\pi }$  differ only in the sign
(the phase factor), and therefore are equivalent to each  other, as well as the
values $\pm \pi$ of the parameter $\theta$.

Now we consider the case $q = 1/2$, for which the angular momentum $j = 0$, and
hence there is no angular momentum degeneracy.
However, the presence of the fermionic zero mode $B_{\pm \pi}$ still leads to a
twofold degeneracy of the  fermion-dyon  system  when  the  parameter $\theta =
\pm\pi$,  since  the  fermionic  zero  mode  can  be  either filled or unfilled
\cite{jr_prd_1976}.
Note in this regard that Eq.~(\ref{III:1}) was  obtained  under  the assumption
that the bound fermionic  state with the positive (negative) energy is unfilled
(filled).
For reasons of continuity, it follows  that the value of $\theta = \pi\,(-\pi)$
corresponds to the unfilled (filled) fermionic zero mode.
Then Eq.~(\ref{III:2}) tells us that the  electric  charge  of the fermion-dyon
system with the unfilled (filled) zero mode is $Q_{\text{u}}=-e/2\;(Q_{\text{f}
} = e/2)$.
Hence, the corresponding fermion number is $N_{\text{u}}=-1/2 \;(N_{\text{f}} =
1/2)$, and  is  half-integral,  as  is  the  case  for  the fermion-kink system
\cite{jr_prd_1976}.
We see that $Q_{\text{f}}-Q_{\text{u}} = e$ and $N_{\text{f}}-N_{\text{u}}= 1$,
as it should be.
The  similar  situation  occurs  when  $\left\vert q \right\vert  >  1/2$.
In this case, the angular momentum $j= \left\vert q \right\vert - 1/2 > 0$, and
the multiplicity of the degeneracy is $2 (2j + 1)= 4 \left\vert q \right\vert$.

\section{Bound states of the fermion-dyon system}                \label{sec:IV}

From the previous section,  we  know  that  the  polarization  of the fermionic
vacuum in  the  vicinity   of   the  Abelian  magnetic  monopole  results  in a
localized spheri- cally-symmetric  distribution  of  the electric charge in the
vicinity of the monopole \cite{yamagishi_prd_1983}.
Thus, the magnetic monopole becomes a dyon  with the electric charge defined by
the Witten formula (\ref{III:2}).
The corresponding electric potential (\ref{III:2a}) has  the long-range Coulomb
asymptotics (\ref{III:4b}).
For positive $q$, this  asymptotics  is attractive for fermions (antifermions),
provided that $\theta > 0$ ($\theta < 0$).
Then, it follows from  the  general consideration of \cite{Landau_III} that due
to the long-range  Coulomb  asymptotics (\ref{III:4b}),  there  are an infinite
number of  bound  (anti)fermionic  states  for  any  value  $j$  of the angular
momentum.
We conclude  that  the  induced  electric  charge  (\ref{III:2})  leads  to the
appearance of new bound fermionic states of the fermion-dyon system.
In addition, the induced electric charge leads to a modification of the already
existing bound fermionic state (\ref{II:36}).

\subsection{\label{subsec:IVA}          Bound fermionic states with the minimal
angular momentum $j = \left\vert q \right\vert - 1/2$}

The presence  of  the  electric  charge  leads  to  modification of the reduced
Hamiltonian (\ref{II:24}) which now takes the form
\begin{equation}
H=-i\frac{q}{\left\vert q\right\vert}\tilde{\gamma}_{5}\frac{d}{dr}
+ M \tilde{\beta} + e A_{0}\mathbb{I},                             \label{IV:1}
\end{equation}
where $\mathbb{I}$ is the two-dimensional identity matrix.
Let us  determine  how  the  presence  of  the  electric  potential  $A_{0}$ in
Eq.~(\ref{IV:1})      affects      the       behaviour       of       solutions
(\ref{II:34}) -- (\ref{II:36}) in the vicinity of $r = 0$. 
According to Eq.~(\ref{III:3b}), the electric  potential increases indefinitely
as $r \rightarrow 0$.
Hence, we can neglect the  mass term $M \tilde{\beta}$ in Eq.~(\ref{IV:1}), and
write the short distance asymptotics of the modified solution as
\begin{equation}
\tilde{u}_{k\theta }\left( r\right) \sim \exp \left[ -ie\tilde{\gamma}
_{5}\int\nolimits_{0}^{r}A_{0}\left( s\right) ds\right] u_{k\theta
}\left( r\right),                                                  \label{IV:2}
\end{equation}
where $u_{k\theta}\left(r\right)$  is  solution  (\ref{II:34}) corresponding to
zero electric potential $A_{0}$.
The solutions (\ref{II:35})  and  (\ref{II:36}) are modified similarly.

Using the short  distance asymptotics (\ref{III:3b}) of the electric potential,
one can show that the integral
\begin{equation}
e\int\nolimits_{0}^{r}A_{0}\left( s\right) ds\sim -\frac{\alpha }{\pi}
qM\sin\left(\theta \right) r\left[\ln \left( Mr\right) \right]^{2} \label{IV:3}
\end{equation}
as $r \rightarrow 0$.
It follows that $\underset{r \rightarrow 0}{\lim}  e \int\nolimits_{0}^{r}A_{0}
\left( s \right)ds  =  0$,   and   therefore   the   exponential   function  in
Eq.~(\ref{IV:2}) tends to unity as $r \rightarrow 0$.
This behaviour  of  the  exponential  function  is  a  consequence  of the weak
(logarithmic)  singularity  of  the  electric  potential $A_{0}$ at the origin.

Based on the  above, we conclude  that  the  presence of the electric potential
$A_{0}$ in Eq.~(\ref{IV:1}) does not lead to a  radical change in the solutions
(\ref{II:34}) -- (\ref{II:36}) in the neighbourhood of $r = 0$.
In  particular,  we   conclude   that  $\tilde{u}_{k\,\theta}\left( 0 \right) =
u_{k\, \theta} \left(0\right) \ne 0$,  and  that  similar  relations  hold  for
solutions (\ref{II:35})  and  (\ref{II:36}).
It follows that the  Hermiticity  condition (\ref{II:28}) and Eq.~(\ref{II:33})
remain satisfied even when the potential $A_{0} \ne 0$.

The integral $\int\nolimits_{0}^{r}A_{0}\left(s\right)ds$ vanishes in the limit
of  small  $r$  also  if  the  potential $A_{0} \sim r^{-\epsilon}$,  where the
exponent $\epsilon \in \left(0, 1\right)$.
Therefore, all conclusions of  the previous paragraph remain valid in this case
too.
In  contrast,  the  integral  diverges  at  the  lower limit  when the exponent
$\epsilon \ge 1$.
In  this  case,  the  exponential  function   in   Eq.~(\ref{IV:2})  oscillates
unboundedly and does not tend to any limit as $r \rightarrow 0$.
This behaviour corresponds to the situation  of falling on the centre in QM and
is unacceptable \cite{Landau_III}.

For example, for the Coulomb potential $A_{0}=-Ze/r$, the solution to the Dirac
equation  $H\chi = E\chi$,  where  $H$  is  given  by  Eq.~(\ref{IV:1}), can be
expressed  in  terms  of  the  confluent Heun   function  \cite{DLMF}  and  its
derivative.
The short distance asymptotics of this solution is
\begin{equation}
\chi(r) \sim \left(
\begin{array}{c}
c_{1}e^{iZ\alpha\ln(Mr)}+c_{2}e^{-iZ\alpha\ln(Mr)} \\
c_{1}e^{iZ\alpha\ln(Mr)}-c_{2}e^{-iZ\alpha\ln(Mr)}
\end{array}
\right),                                                          \label{IV:3b}
\end{equation}
where $c_{1}$ and $c_{2}$ are arbitrary complex constants.
We see  that  as  $r \rightarrow 0$, the  solution  oscillates  with increasing
frequency, which corresponds  to  falling  of the fermion on the dyon's centre.
As $r \rightarrow 0$, the asymptotics  of  the ratio of the radial functions is
\begin{equation}
\frac{f\left( r\right) }{g\left( r\right)}\sim 1 - \frac{2c_{2}}
{c_{2}-c_{1}e^{2iZ\alpha \ln (r)}}.                               \label{IV:3c}
\end{equation}
Except for the two degenerate  cases  $c_{1} = 0$   and  $c_{2} = 0$, the ratio
(\ref{IV:3c}) does not tend to a certain limit as $r \rightarrow 0$.
In the two degenerate cases, this ratio is $\mp 1$ and is real.
We see that the  asymptotics (\ref{IV:3b})  is  incompatible  with the boundary
condition (\ref{II:33}).

It was shown in  Section~\ref{sec:II}  that  $CP$  inversion  leaves the vector
potential of the monopole unchanged.
Due to  this, $CP$  is  a symmetry  of  the  fermion-monopole  system, provided
that its angular momentum $j \ge \left\vert q \right\vert + 1/2$.
At the same time, the  boundary  condition (\ref{II:33}) is not invariant under
$CP$, which changes sign of the parameter $\theta$.
Hence, $CP$ is not a  symmetry  of  the fermion-monopole system with the lowest
angular momentum $j = \left\vert q \right\vert - 1/2$.

The situation changes  when  we consider  the  polarization  of  the  fermionic
vacuum, which is equivalent to the transformation  of the monopole into a dyon.
Since the electric charge  of  the  dyon  is  nonzero,  it  possesses  a radial
electric  field  $E_{r} = -\partial_{r}A_{0}$  corresponding  to  the  electric
potential $A_{0}$.
Unlike the monopole's magnetic  potential   $\mathbf{A}$,  the  dyon's electric
potential $A_{0}$ changes sign under $CP$ inversion.
Hence, $CP$ transforms  the  original  dyon  into another one with the opposite
electric charge.
As a consequence, $CP$ cannot  be  a  symmetry  of the fermion-dyon system even
when its angular momentum $j \ge \left\vert q \right\vert + 1/2$.

Without considering the electric charge,  the fermion-monopole system possesses
the single bound  (anti)fermi- onic  state  (\ref{II:36})  existing  due to the
nontrivial boundary condition (\ref{II:33}).
Accounting for electric charge  leads  to  corrections  to the wave function of
the bound state (\ref{II:36}) and to its energy.
These corrections   can   be   calculated   either   numerically  or  using the
perturbation theory.
In the latter case, the first order correction to the energy is
\begin{equation}
\Delta E^{(1)}=\left(B_{\theta},\Delta HB_{\theta}\right),        \label{IV:3d}
\end{equation}
where the perturbation $\Delta H = e A_{0} \mathbb{I}$.
Using  Eqs.~(\ref{II:36})  and  (\ref{III:2a}),  and  applying  analytical  and
approximate methods, we obtain an approximate   analytical   expression for the
first-order energy correction
\begin{eqnarray}
\Delta E^{(1)} &\approx &-2Mq\frac{\alpha }{\pi}\theta \cos (\theta
)\ln \left[ -\cos (\theta )\right]                                  \nonumber
 \\
&&+\frac{Mq}{12}\frac{\alpha }{\pi }\text{sgn}\left( \theta \right) \cos
(\theta )\left\{ 12\pi (1+\ln(2))\right.                            \nonumber
 \\
&&\left.+12(\ln(4)-5)\cos(\theta )-\pi \cos^{2}(\theta)\right\}, \label{IV:4}
\end{eqnarray}
where $\left\vert \theta \right\vert \in \left[ \pi /2,\pi \right]$.
We see that the correction to the energy $E= M \sin(\theta)$ of the bound state
(\ref{II:36}) is small, since the fine structure constant $\alpha \approx 1/137
\ll 1$.
Furthermore,  it  changes  its  sign  under  $CP$  inversion  at  which $\theta
\rightarrow -\theta$.

\begin{figure}[tbp]
\includegraphics[width=0.5\textwidth]{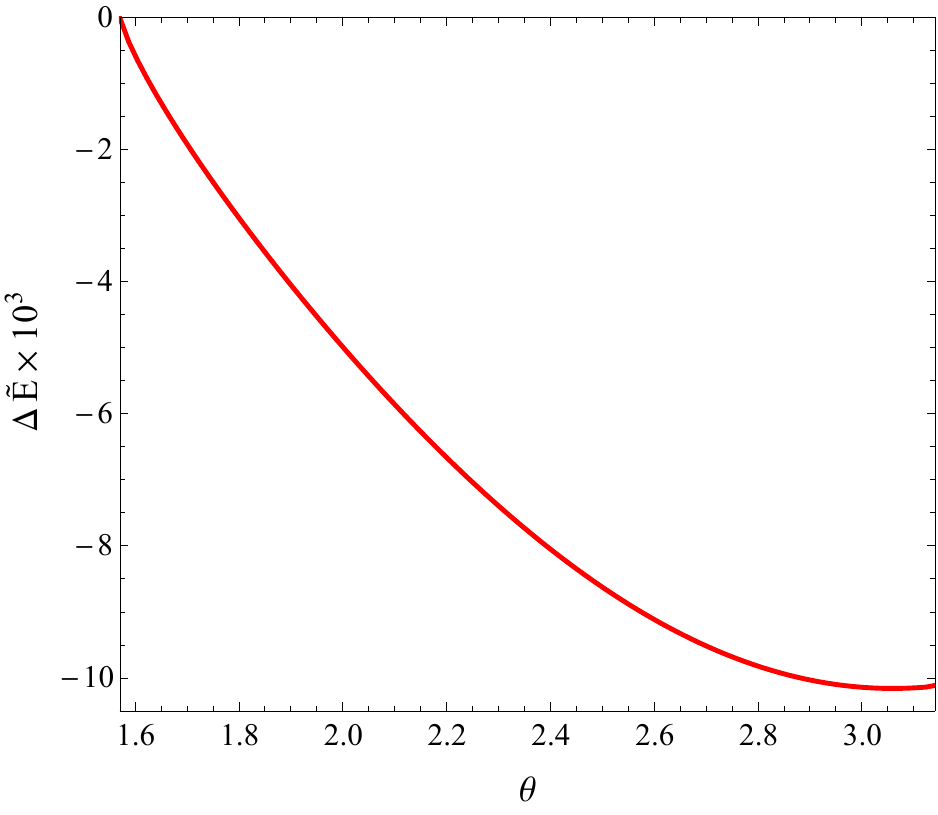}
\caption{\label{fig2}    Dependence   of   the   correction  $\Delta\tilde{E} =
\Delta E/M$ to  the  energy  of  bound  state  (\ref{II:36})  on  the parameter
$\theta$.  The correction corresponds to the parameters $\alpha = 1/137$ and $q
= 1/2$}
\end{figure}

Figure~\ref{fig2} shows  the  dependence of the correction to the energy of the
bound state (\ref{II:36}) on the parameter $\theta$.
The correction was found  by numerically solving  the first-order system $H\chi
=E\chi$ with the perturbed Hamiltonian (\ref{IV:1}).
Note that the  curve  $\Delta E^{(1)}$  obtained  using  Eqs.~(\ref{IV:3d}) and
(\ref{IV:4}) is visually indistinguishable from that  shown in Fig.~\ref{fig2}.
This is because  the  difference $\Delta E - \Delta E^{(1)}$ is proportional to
$\alpha^{2}$, where  the  fine structure constant $\alpha \approx 1/137 \ll 1$.
It follows from  Fig.~\ref{fig2}  that  the  presence  of  the  electric charge
increases the binding energy  of  the fermion-dyon system   because  the  signs
of the electric charges of the dyon and fermion are opposite.
As a result,  the curve $E(\theta) = M \sin(\theta) + \Delta E(\theta)$ crosses
zero not at the point $\theta=\pi$ as in the unperturbed case $\Delta E(\theta)
= 0$, but at the nearby point $\theta \approx \pi - 2.76 q \alpha$.

Eq.~(\ref{IV:4}) tells us that  in  the  neighbourhood of $\theta = \pi/2$, the
correction $\Delta E^{(1)}$ can be written as
\begin{equation}
\Delta E^{(1)} \approx \alpha q M \delta \left(\ln \left(\delta/2\right)
-1 \right),                                                       \label{IV:4a}
\end{equation}
where the parameter $\delta = \theta - \pi/2 >0$.
On the other  hand,  the   binding   energy   of  the  unperturbed  bound state
(\ref{II:36}) is
\begin{equation}
E^{\left(\text{b}\right)}_{\text{u}}=M\left(\sin \left(\pi/2+\delta\right)
-1\right) \approx -M\delta ^{2}/2.                                \label{IV:4b}
\end{equation}
From Eqs.~(\ref{IV:4a}) and  (\ref{IV:4b}),   it   follows  that the correction
(\ref{IV:4a})  exceeds  the  unperturbed  binding  energy  (\ref{IV:4b}) in the
magnitude   provided   that   $\pi/2  <  \theta  \lesssim  \pi/2  -  2 \alpha q
\ln \left(\alpha \right)$.
We see that in this small neighbourhood  of  $\theta$,  the role of the  dyon's
electric  charge  becomes   comparable   to  that  of  the  boundary  condition
(\ref{II:33}).

Besides the corrections to the already existing  bound state (\ref{II:36}), the
electric charge of the  dyon  gives  rise  to new bound states with the angular
momentum $j = \left\vert q \right\vert - 1/2$.
Indeed, at large distances,  the  electric  potential $A_{0}$ is attractive and
has long-range Coulomb asymptotics (\ref{III:4b}).
In accordance with the considerations  of \cite{Landau_III}, this should result
in the existence  of  an  infinite  number  of  bound fermionic (antifermionic)
states for positive (negative) $\theta$.
In contrast  to the state (\ref{II:36}), these states are  loosely bound due to
the smallness of the fine structure constant $\alpha \approx 1/137$.
Furthermore, these states exist in the range of $0< \left\vert\theta\right\vert
\le\pi$, whereas the bound state (\ref{II:36}) exists only if $\pi/2<\left\vert
\theta \right\vert \le \pi$.

To find the wave functions and  energies  of  these bound states, we must solve
an eigenvalue problem.
The feature of our case is that the radial wave functions $f(r)$ and $g(r)$ are
nonzero at $r = 0$ because of the boundary condition (\ref{II:33}).
This  inhomogeneous  boundary   condition   complicates   the  solution  of the
eigenvalue problem.
To avoid this, we introduce the combination $F\left(r\right) = f\left(r\right)-
i \tan\left[ \theta /2 + \pi /4 \right] g\left(r \right)$ for which $F(0) = 0$.
We define the dimensionless radial function
\begin{equation}
\phi \left( r\right) =\left[E-M\sin \left(\theta\right)-eA_{0}\left(
r\right) \right] ^{-1/2}F\left( r\right),                         \label{IV:4c}
\end{equation}
which satisfies  the  homogeneous boundary  condition $\phi\left(0\right) = 0$.
Then, we  reduce  the  first-order  system  $H\chi  = E\chi$ with the perturbed
Hamiltonian (\ref{IV:1}) to the  second-order equation for the  radial function
$\phi \left( r \right)$
\begin{equation}
\phi^{\prime\prime} + \left[-\varkappa^{2}-V\right]\phi = 0,  \label{IV:4d}
\end{equation}
where
\begin{eqnarray}
V &=&2eA_{0}E-\frac{eMA_{0}^{\prime }\cos (\theta )}{E-M\sin (\theta
)-eA_{0}}                                                           \nonumber
 \\
&&+\frac{3}{4}\frac{e^{2}A_{0}^{\prime 2}}{(E-M\sin(\theta)-eA_{0})^{2}}
                                                                    \nonumber
 \\
&&+\frac{1}{2}\frac{eA_{0}^{\prime \prime }}{E-M\sin (\theta)-eA_{0}}
-e^{2}A_{0}^{2}                                                    \label{IV:4e}
\end{eqnarray}
and $\varkappa^{2}=M^{2}-E^{2}$.

Eq.~(\ref{IV:4d})  formally  coincides  with  a one-dimensional Schr\"{o}dinger
equation with the energy $-\kappa^{2}$  and  the  potential $V$  defined on the
semi-infinite interval $r \ge 0$.
The only  difference  is  that  the  potential  $V$  depends  on  the fermion's
energy $E$.
A characteristic  feature   of    the   potential   $V$  is  the  absence  of a
centrifugal barrier.
As  a  consequence,  the   potential   $V  \sim   1/(r^{2}\ln(M r))$   at short
distances.
At large  distances,  the  potential  $V$  possesses  a  long-range asymptotics
$\propto r^{-1}$.
The nonanalytical   behaviour   of   $V$    at  short  distances  excludes  the
possibility of  an  analytical  solution  of  eigenvalue problem (\ref{IV:4d}).
To solve the eigenvalue problem, we  use  numerical  methods implemented in the
{\sc{Mathematica}} software package \cite{Mathematica}.

\begin{figure}[tbp]
\includegraphics[width=0.5\textwidth]{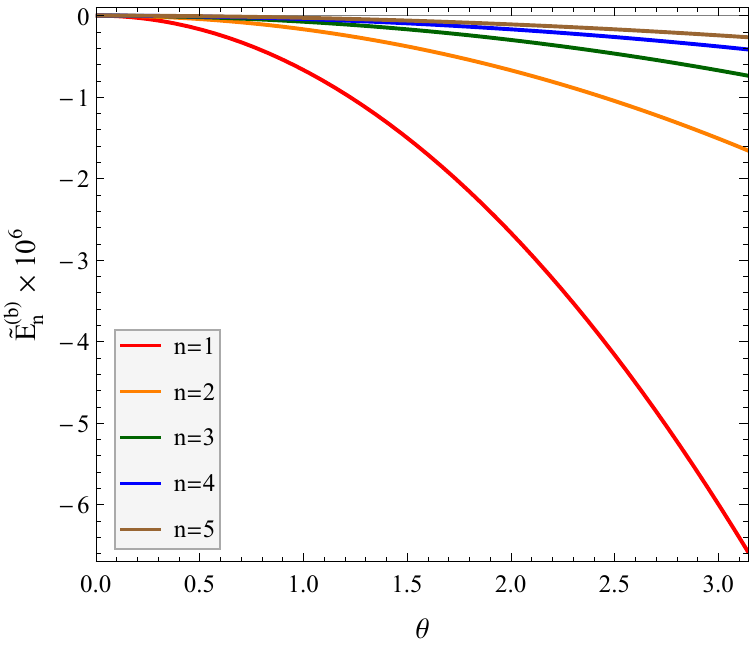}
\caption{\label{fig3}            Dependence of the dimensionless binding energy
$\tilde{E}^{(\text{b})}_{n} = (E_{n} - M)/M$  of the first  five loosely  bound
fermionic states of the third type  on the  parameter  $\theta$.     The curves
correspond to the parameters $\alpha = 1/137$ and $q = 1/2$}
\end{figure}

Figure~\ref{fig3} shows the dependence of the binding energy $E^{(\text{b})}_{n
}=E_{n} - M$ of the  first  five  bound  fermionic  states of the third type on
the parameter $\theta$.
We consider the most realistic case of $\alpha = 1/137$ and $q = 1/2$.
It was found numerically that all  the  curves  $E^{(\text{b})}_{n}(\theta)$ in
Fig.~\ref{fig3} can be described with high accuracy by the quadratic dependence
\begin{equation}
E_{n}^{\left(\text{b}\right)}(\theta)\approx-a_{n}\theta^{2},     \label{IV:4f}
\end{equation}
where the radial quantum number $n=1,2,\ldots$ is the number of zeros of either
of the two radial functions $f$ and  $g$,  and  the  coefficient $a_{n} \approx
6.5847 \times 10^{-6} n^{-2}$.
We see,  in Fig.~\ref{fig3}, that the binding energies $E^{(\text{b})}_{n}$ are
much less than the fermion mass $M = 1$, and therefore the fermions are loosely
bound and nonrelativistic.
This is because the fine  structure  constant  $\alpha  = 1/137 \ll 1$, and the
magnitude of electric charge of  the  dyon  does  not  exceed  $e/2$  when $q =
1/2$.
On the contrary, the binding  energy  of  state  (\ref{II:36}) is approximately
$M(\sin(\theta)-1)$, and, in the general case, this is of the same order as the
fermion mass $M$.
Hence,  in  this  case,   the   fermion  is  tightly  bound  and  relativistic.
The  reason  is  that  the  existence  of  the bound  state  (\ref{II:36}) only
depends  on  the  boundary  condition (\ref{II:33}) and  does not depend on the
electric charge of the dyon.

\subsection{\label{subsec:IVB}          Bound fermionic states with the angular
momentum $j \ge \left\vert q \right\vert + 1/2$}

We  now  consider   the  existence  of  bound  fermionic  states  with  angular
momentum $j \ge \left\vert q \right\vert + 1/2$.
It was shown in \cite{kzm1_prd_1977}  that  there  are  two  types of fermionic
states in this case.
The states of the first type have the form
\begin{equation}
\psi _{j m}^{\left( 1\right) }(t,\mathbf{x})=\frac{1}{r}
\begin{bmatrix}
f\left( r\right) \xi _{j m}^{\left( 1\right) }\left( \vartheta ,\varphi
\right)  \\
g\left( r\right) \xi _{j m}^{\left( 2\right) }\left( \vartheta ,\varphi
\right)
\end{bmatrix}
e^{-iEt},                                                          \label{IV:5}
\end{equation}
where the  two-component  angular  momentum  eigensections  $\xi_{j m}^{\left(1
\right) }\left(\vartheta ,\varphi \right)$   and  $\xi _{j m}^{\left( 2\right)}
\left(\vartheta, \varphi \right)$ were defined in \cite{kzm1_prd_1977}.
The states  of  the  second  type  are  obtained  from  Eq.~(\ref{IV:5}) by the
permutation  $\xi_{j m}^{\left( 1 \right)}  \leftrightarrow  \xi_{j m}^{\left(2
\right)}$.
In the following, we shall limit  our  discussion  to states of the first type,
since the properties of the two types of states are close enough.

Substituting Eq.~(\ref{IV:5}) into the  Dirac  equation (\ref{II:1}), we obtain
the system of first-order equations for the radial wave functions
\begin{eqnarray}
i\left(\partial_{r}-\mu r^{-1}\right)f-\left(M + E - eA_{0}\right)g & = &0,
                                                                  \label{IV:6a}
 \\
i\left(\partial_{r}+\mu r^{-1}\right)g+\left(M-E+eA_{0}\right)f & = &0,
                                                                  \label{IV:6b}
\end{eqnarray}
where the parameter $\mu = \bigl[\left(j + 1/2\right)^{2} - q^{2}\bigr]^{1/2}$.
We define dimensionless  radial  functions  $u(r)$  and $v(r)$ as
\begin{subequations}                                               \label{IV:7}
\begin{eqnarray}
u\left( r\right)  &=&\left[E+M-eA_{0}\left(r\right)\right]^{-1/2}f\left(
r\right),                                                         \label{IV:7a}
  \\
v\left( r\right)  &=&\left[E-M-eA_{0}\left(r\right)\right]^{-1/2}g\left(
r\right).                                                         \label{IV:7b}
\end{eqnarray}
\end{subequations}
Now  we turn to the new radial functions  and  reduce the resulting first-order
system  to a second-order equation for one of the radial functions
\begin{equation}
u^{\prime \prime }+\left[ -\varkappa ^{2}-U\right] u=0,            \label{IV:8}
\end{equation}
where 
\begin{eqnarray}
U &=&2eA_{0}E-\frac{\mu\left(1-\mu\right)}{r^{2}}+\frac{e\mu A_{0}^{\prime}}
{E+M-eA_{0}}\frac{1}{r}                                            \label{IV:9}
 \\
&&+\frac{3}{4}\frac{e^{2}A_{0}^{\prime 2}}{\left(E+M-eA_{0}\right)^{2}}+
\frac{1}{2}\frac{eA_{0}^{\prime\prime}}{E+M-eA_{0}}-e^{2}A_{0}^{2}  \nonumber
\end{eqnarray}
and  $\varkappa^{2} = M^{2} - E^{2}$.

Similarly to  Eq.~(\ref{IV:4d}), Eq.~(\ref{IV:8})  formally  coincides  with  a
one-dimensional Schr\"{o}dinger  equation  with   energy  $-\kappa^{2}$  and an
energy-dependent potential $U$ defined on the semi-infinite interval $r \ge 0$.
Using  Eqs.~(\ref{III:4b})  and  (\ref{IV:9}),  we  obtain  the  large-distance
asymptotics of the potential  $U$
\begin{eqnarray}
U &\sim &-2E\frac{z\alpha}{r}+\frac{\mu\left(\mu-1\right) }{r^{2}}  \nonumber
 \\
&&-\frac{z^{2}\alpha^{2}}{r^{2}}+\frac{\left(\mu-1\right)}{E+M}
\frac{z \alpha}{r^{3}}+O\left[ \frac{\alpha^{2}}{r^{4}}\right],   \label{IV:10}
\end{eqnarray}
where the parameter $z = q\theta/\pi$.
Similarly,   using   Eqs.~(\ref{III:3b})   and  (\ref{IV:9}),   we   obtain the
short-distance asymptotics of  $U$
\begin{equation}
U\sim \frac{\mu \left( \mu -1\right) }{r^{2}}+\frac{1-2\mu }{\ln \left(
Mr\right)}\frac{1}{r^{2}}+O\left[\alpha^{2}\ln^{4}\left(Mr\right)
\right].                                                          \label{IV:11}
\end{equation}
Eq.~(\ref{IV:10}) tells us  that  at  large distances, the potential $U \propto
r^{-1}$  in  leading  order  in $r$, and  is  attracting  (repelling)  provided
that the sign of $z E$ is positive (negative).
From Eq.~(\ref{IV:11}), it follows  that  due  to  the centrifugal barrier, the
potential $U \propto r^{-2}$, and  is repelling at small distances.
Because  of  a   mild  (logarithmic)  singularity  in  Eq.~(\ref{III:3b}),  the
contribution  from  the   electric  potential  $A_{0}$  to  the  short-distance
asymptotics (\ref{IV:11}) is suppressed logarithmically compared to the leading
term.
Furthermore, in Eq.~(\ref{IV:11}),  neither  the  leading  nor subleading terms
depend on $\alpha$ or $z$.

A characteristic feature of the potential in Eq.~(\ref{IV:9}) is that it can be
approximated by a relatively simple expression,
\begin{equation}
\bar{U}=-2E\frac{ze^{2}}{r}+\frac{\mu\left(\mu-1\right)}{r^{2}}.  \label{IV:12}
\end{equation}
For  $r  >  M^{-1}$  ($r  <  M^{-1}$),  the  relative  error  of  approximation
(\ref{IV:12}) does not exceed $10^{-4}$  $(10^{-2})$, provided  that  the  fine
structure constant $\alpha \lesssim 1/137$.
We can therefore  use  the  approximate potential (\ref{IV:12}) to describe the
bound states of the fermion-dyon system.
Thus, we shall consider the approximate equation
\begin{equation}
\bar{u}^{\prime\prime} + \bigl[-\varkappa^{2}
-\bar{U} \bigr]\bar{u} = 0,                                       \label{IV:14}
\end{equation}
where we use the bar to denote the approximate quantities.
Based on the general  considerations  of  \cite{Landau_III},  we  can  say that
Eq.~(\ref{IV:8}) has an infinite number of bound  states  for  any $j \ge \left
\vert q \right\vert + 1/2$, provided that $z E > 0$.

Indeed, using standard methods of the theory of differential equations, we find
that for each $j \ge \left\vert q \right\vert+1/2$, Eq.~(\ref{IV:14}) possesses
an infinite number of normalizable solutions
\begin{equation}
\bar{u}_{n j}=\mathcal{N}_{n j}(2\varkappa_{n j}r)^{\mu}
\,e^{-\varkappa_{n j}r}L_{n}^{2\mu -1}(2\varkappa_{n j}r),        \label{IV:15}
\end{equation}
where $L_{n}^{2\mu-1}$ is the generalized Laguerre polynomial  with  the radial
quantum number $n=0,1,2,\ldots$, $\mathcal{N}_{n j}$ is a normalization factor,
and the parameter $\varkappa_{n j}=\bigl[M^{2}-\bar{E}_{nj}^{2}\bigr]^{1/2}$.
The energies of the bound (anti)fermionic states are
\begin{equation}
\bar{E}_{n j} = \sigma M\left[1+\frac{z^{2}\alpha^{2}}{\left(n+\mu \right)
^{2}}\right]^{-1/2},                                              \label{IV:17}
\end{equation}
where $\sigma = \text{sgn}\left(z\right)$ and $\mu = \bigl[\left(j+1/2\right)^{
2}-q^{2}\bigr]^{1/2}$.

In the transition from $\bar{u}_{n j} $ to $ \bar{f}_{n j}$, we can neglect the
term $e A_{0}(r)$ under the radical sign in Eq.~(\ref{IV:7a}).
This is because, according to Eq.~(\ref{IV:15}), noticeable  changes  of $\bar{
u}(r)$ occur on the scale on the order of $\varkappa_{n j}^{-1} \approx \left(n
+  \mu  \right)\times\\(M  \alpha  z)^{-1}$,   whereas  those  of  the  radical
$\sqrt{E + M + eA_{0}}$   occur  on  the  much  smaller  scale  on the order of
$M^{-1}$.
Due to this, the function $\bar{u}_{n j}(r)$ is  still close to zero, while the
radical reaches the limit of $\sqrt{E + M}$.
It follows   that   the   region  $r\lesssim O\left[ M^{-1} \right]$   makes no
appreciable contribution to observables, and therefore  the exact form there of
the radical is unimportant.

Thus, we conclude  that  the  radial  wave  function  $f_{n j}$  can be written
approximately as
\begin{equation}
\bar{f}_{n j}=\bigl[\bar{E}_{n j}+M\bigr]^{1/2}\bar{u}_{n j}.
                                                                  \label{IV:18}
\end{equation}
Substituting Eq.~(\ref{IV:18}) into Eq.~(\ref{IV:6a}), we obtain an approximate
expression for the radial wave function $g_{n j}$
\begin{equation}
\bar{g}_{n j}= i \bigl[\bar{E}_{n j}+M\bigr]^{-1}
\bigl[\bar{f}_{n j}^{\prime}-\mu\bar{f}_{n j}/r \bigr].       \label{IV:19}
\end{equation}
Using the normalization condition $\int\nolimits_{0}^{\infty }\bigl( \left\vert
\bar{f}_{nj}\right\vert^{2} +\left\vert \bar{g}_{nj}\right\vert^{2}\bigr)dr\\=1$,
we find that the normalization factor
\begin{equation}
\mathcal{N}_{n j}=\biggl[ 1+\frac{\left( n+\mu \right) ^{2}}{
z^{2}\alpha^{2}}\biggr]^{-1/4}\left[\frac{n!}{2\left(n+\mu\right)
\mathrm{\Gamma}(n+2\mu )}\right] ^{1/2}.                          \label{IV:20}
\end{equation}

In addition to the bound states of  the  first  type,  the  fermion-dyon system
also possesses an infinite number of bound states of the second type.
The energies of these bound  states  are obtained from Eq.~(\ref{IV:17}) by the
substitution $n \rightarrow n + 1$.
As a result, we have the  relation $\bar{E}_{n + 1\text{\/}j}^{\left( 1\right)}
= \bar{E}_{n\text{\/}j}^{\left(2\right)}$,  from  which  it  follows  that  all
bound levels  are  doubly  degenerate  except  for the $\bar{E}_{0\text{\/}j}^{
\left(1\right)}$ levels.
Note, however, that this twofold degeneracy takes place only   within  the used
approximation, and is removed when corrections are taken into account.

Let us estimate the accuracy of the used approximation.
To do this, we must  estimate  the  mean  value  of  the  perturbation operator
$\Delta U = U-\bar{U}$ in the state $\bar{\psi}^{(1,2)}_{n j}$.
Using  Eqs.~(\ref{IV:9}) -- (\ref{IV:20})   and   approximate   and  analytical
methods, we can  show  that  $\left\langle \Delta U \right \rangle_{nj} \propto
M z^{4} \alpha^{4}$.
Using the results obtained  for  the  hydrogen  atom,  we  can  assume that the
radiative  corrections  $\Delta E_{\text{rad}}$  to  the  bound (anti)fermionic
levels are $\propto M z^{4} \alpha^{5} \ln\left(1/\alpha\right)$.
In view of the above,  we  conclude  that Eq.~(\ref{IV:17}) is only valid up to
terms of order $z^{2}\alpha^{2}$, and hence can be rewritten as
\begin{equation}
\bar{E}_{nj}=\sigma M\left[1-\frac{z^{2}\alpha^{2}}
{2\left( n+\mu \right)^{2}}\right].                               \label{IV:21}
\end{equation}
The relative error of Eq.~(\ref{IV:21}) is on  the  order of $z^{2}\alpha^{2}$.
We see that in the most interesting case of $\left\vert q \right\vert =1/2$ and
$\alpha \approx 1/137$, this error is on the order of 0.005\textdiv 0.01\%.
The difference between $E_{nj}$ obtained by numerical methods and $\bar{E}_{nj
}$ of Eq.~(\ref{IV:21}) is in agreement with this estimate.

Eqs.~(\ref{IV:18})  and  (\ref{IV:19})  were  obtained  under  the  assumptions
$E > 0$ and $\sigma = \text{sgn}\left(z\right) = 1$.
The approximate antifermionc wave functions are obtained from Eqs.~(\ref{IV:18})
and (\ref{IV:19}) via the substitution ($CP$ inversion)
\begin{equation}
\bar{f}_{nj}\rightarrow -i\bar{g}_{nj},\,
\bar{g}_{nj}\rightarrow  i\bar{f}_{nj},\,
\bar{E}_{nj}\rightarrow -\bar{E}_{nj}.                            \label{IV:22}
\end{equation}
Note that substitution (\ref{IV:22}) converts  fermionic  states  of  the first
(second) type into antifermionic states of the second (first) type.

The approximated energy spectrum (\ref{IV:17}) is hydrogen-like.
Formally, the spectrum of the hydrogen  atom is obtained from Eq.~(\ref{IV:17})
by the substitution $z \rightarrow 1,\, q \rightarrow \alpha$.
However, the angular momentum $j$  is  half-integral   for  the  hydrogen atom,
whereas, it is an integer (half-integer)  for the fermion-dyon system, provided
that $q$ is a half-integer (integer).
The  approximated  spectrum (\ref{IV:17}) is similar to the exact spectrum of a
fermion-dyon    system    obtained    in   \cite{zhang_prd_1986,zhang_prd_1989,
zhang_jmp_1990,zhang_prd_1990} for the case of the Coulomb  electric  potential
and nonminimal angular momenta $j \ge \left\vert q \right\vert + 1/2$.
This spectrum  can be obtained   from  Eq.~(\ref{IV:17})  by  the  substitution
$z  \rightarrow  Z_{D},\,  q^{2} \rightarrow q^{2} + (Z_{D} \alpha)^{2}$, where
$Z_{D}$ is  an  integer, and $Z_{D}e$ is the dyon's charge.

The energy levels of the  Coulomb  fermion-dyon  system  are doubly degenerate,
satisfying the relation $E^{(1)}_{nj} = E^{(2)}_{nj}$.
Unlike the twofold  degeneracy  of  the  fermion-dyon  system  considered here,
the twofold  degeneracy  of  the Coulomb fermion-dyon  system  is  exact at the
quantum mechanical level, but it  is probably removed by radiative corrections.
This exact  quantum  mechanical  twofold  degeneracy  is  a  consequence of the
Coulomb  electric   potential   used   in  \cite{zhang_prd_1986,zhang_prd_1989,
zhang_jmp_1990,zhang_prd_1990}.
In our case, the electric potential is  essentially non-Coulomb for $r \lesssim
M^{-1}$, and due to  this,  the  twofold  degeneracy  of  the  energy levels is
removed already at the quantum mechanical level.

The  rich  structure  of  the  energy   levels   of   the  fermion-dyon  system
considered here  implies  the  possibility  of  numerous  radiative transitions
between them.
It is similar with the  hydrogen  atom,  but  there is an important difference:
because of the  violation  of  parity,  the  electric  dipole  transitions with
$\Delta j = 0$  are  allowed  for  a  fermion-dyon system \cite{shnir_jpg_1988,
zhang_prd_1990}, as well  as  $\Delta j  =  \pm 1$  transitions, unlike for the
hydrogen  atom,  where   parity  conservation  strictly  forbids $\Delta j = 0$
transitions and allows only  $\Delta j  =  \pm 1$ transitions.

Another difference from  the  hydrogen  atom  is the tightly bound state of the
fermion-dyon system.
In the general case, the  binding  energy  of  this  state exceeds those of the
loosely bound states by orders of magnitude.
As a result, the energy of a  photon  emitted  by  a  transition from a loosely
bound state to the  tightly  bound  state  exceeds  by  orders of magnitude the
energies of photons emitted by transitions between loosely bound states.
In  contrast,  there  is  no  analog  of   the  tightly  bound   state  of  the
fermion-dyon  system  in   the   hydrogen   atom,  and   therefore  the maximum
possible energy of an emitted photon  is  of the same order of magnitude as the
energies of some other emitted photons.

\section{Electric dipole moments of the bound fermionic states}   \label{sec:V}

It  was  shown  in   Section~\ref{sec:II}  that   $P$   transformation  changes
sign  of  the  magnetic  charge  of  the  monopole,  and  therefore cannot be a
symmetry of the fermion-monopole system.
In  particular,  the  bound  fermionic  states  are  not  eigenstates   of  $P$
transformation.
This makes it possible  for  the  bound  states  of  the fermion-dyon system to
have a nonzero electric dipole moment.
On the  contrary,  if  a   bound    fermionic   state   is   an  eigenstate  of
$P$, as is the  case  for  the hydrogen atom,  then it cannot possess a nonzero
electric dipole moment.
This is because the electric dipole moment operator $\mathbf{d} = e \mathbf{r}$
changes sign  under  $P$  transformation,  whereas  the  bound  fermionic state
remains unchanged modulo a phase factor.

It follows  from  the  conservation   of   angular   momentum   that   the only
nonvanishing component  of  the  operator  $\mathbf{d}$ in a state $\psi_{j m}$
is the $z$ component
\begin{equation}
\left\langle d_{z}\right\rangle_{j\,m} = e\int \psi_{jm}^{\dag}
z \psi_{j m} d^{3}x.                                                \label{V:1}
\end{equation}
It was shown in Section~\ref{sec:IV}  that  the induced  electric charge of the
dyon results  in  the  existence  of  an  infinite  number  of  bound fermionic
states for each admissible value $j$ of the angular momentum.
Each bound  state  of  the  fermion-dyon  system   possesses a nonzero electric
dipole moment, provided that $\left\vert q \right\vert > 1/2$.
We shall  consider  the  cases  of  bound  fermionic  states  with  the minimal
($j = \left\vert q \right\vert -1/2$) and nonminimal ($j \ge \left\vert q\right
\vert + 1/2$) angular momentum separately.

\subsection{\label{subsec:VA}          Electric  dipole  moments  of  the bound
fermionic states with the angular momentum $j=\left\vert q \right\vert - 1/2$}

In this case, the  bound  fermionic  states  are  of  the  third  type, and the
expression of the electric dipole moment takes the form
\begin{equation}
\left\langle d_{z}\right\rangle_{jm} = e \!  \int \! \eta_{jm}^{\dag}
\eta_{jm}\cos(\vartheta)d\Omega \int\nolimits_{0}^{\infty}\! r
\bigl(\left\vert f \right\vert^{2} + \left\vert g \right\vert^{2}\bigr)dr.
                                                                    \label{V:2}
\end{equation}
It was shown in \cite{kzm3_prd_1977} that the kinematic factor
\begin{equation}
\int \eta _{jm}^{\dag }\eta _{jm}\cos \vartheta \,d\Omega = -\text{sgn}
\left( q\right) \frac{m}{j + 1},                                    \label{V:3}
\end{equation}
where the angular  momentum  $j = \left\vert q \right\vert - 1/2$  and  the $z$
projection $m\in \left[-\left\vert q \right\vert+1/2, \left\vert q \right\vert-
1/2\right]$.
Note that  the  kinematic  factor  vanishes  in  the  important  case  of  $q =
\pm 1/2$, which corresponds to the  elementary (the winding number $n = \pm 1$)
magnetic (anti)monopole.

Substituting the  radial wave functions  from Eq.~(\ref{II:36}) into the second
integral in Eq.~(\ref{V:2}), and using Eq.~(\ref{V:3}), we obtain an analytical
expression for  the  electric  dipole  moment  of  the  unperturbed bound state
(\ref{II:36})
\begin{equation}
\left\langle d_{z}\right\rangle _{jm} =
-\text{sgn}\left( q\right) \frac{m}{j+1}
\frac{e}{2 M \left\vert \cos \theta \right\vert }.                  \label{V:4}
\end{equation}
Eq.~(\ref{V:4}) tells us that the  electric  dipole  moment  of the unperturbed
state (\ref{II:36}) increases indefinitely  $\propto (\pi/2 - \left\vert \theta
\right\vert)^{-1}$ as $\left\vert \theta \right\vert \rightarrow \pi/2$.
The reason is that  in  this  limit,  the  state  (\ref{II:36}) becomes loosely
bound, and therefore the fermion is weakly localized.
In contrast, the state (\ref{II:36}) is  tightly  bound when $\left\vert \theta
\right\vert$ is in the neighbourhood of $\pi$.
In this case, the radial integral  in  Eq.~(\ref{V:2})  is  on the order of the
Compton wavelength of the fermion, and  the  dipole  moment reaches its minimum
values.

\begin{figure}[tbp]
\includegraphics[width=0.5\textwidth]{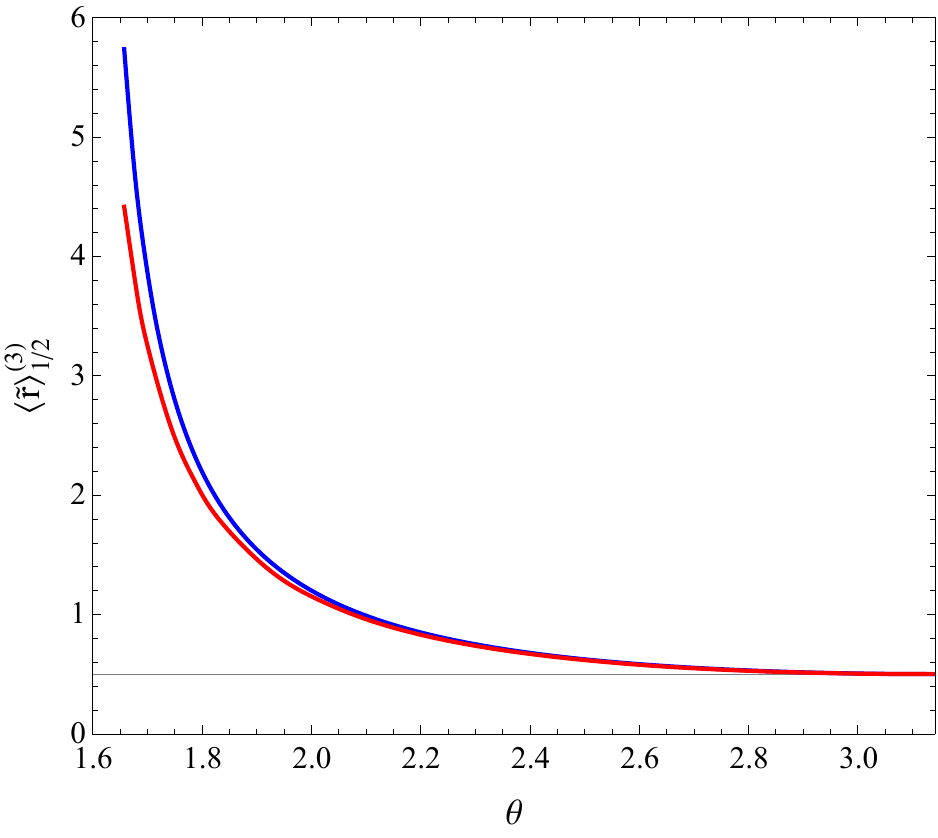}
\caption{\label{fig4}    Dependence  of  the  dimensionless  mean radius of the
fermion spatial distribution $\langle \tilde{r} \rangle^{(3)}_{1/2} = M \langle
r \rangle^{(3)}_{1/2}$ on the parameter $\theta$.   The  red  (blue)  curve  is
obtained with (without) taking  into  account  the electric charge of the dyon.
The curves correspond to the parameters $\alpha = 1/137$ and $q = 1$}
\end{figure}

Eq.~(\ref{V:4}) is obtained without  accounting  for the electric charge of the
dyon.
To account for it,  we  must  solve  the  Dirac  equation  with  the  perturbed
Hamiltonian (\ref{IV:1}) and substitute the obtained radial wave functions into
Eq.~(\ref{V:2}).
To solve the Dirac equation, we use numerical methods of the {\sc{Mathematica}}
software package \cite{Mathematica}.

It follows from Eq.~(\ref{V:2}) that the electric dipole moment is proportional
to the mean radius of the fermion  spatial distribution $\langle r \rangle^{(3)
}_{j}  =  \int  \nolimits_{0}^{\infty}  r  (\left\vert  f_{j}  \right\vert^{2}+
\left\vert g_{j}\right\vert^{2})dr$, where the type and angular momentum of the
fermionic state are indicated.
Figure~\ref{fig4} shows  the  dependence   of   the  dimensionless  mean radius
$\langle\tilde{r} \rangle^{(3)}_{j}  =  M  \langle r \rangle^{(3)}_{j}$  on the
parameter $\theta$ for two cases.
In the first case,  the  contribution  of  the dyon's electric charge was taken
into account, whereas in the second case it was not.
It follows from Eq.~(\ref{V:4})  that  in  the  second  case,  the  mean radius
$\langle \tilde{r} \rangle^{(3)}_{j}=2^{-1} \left\vert \cos\left(\theta \right)
\right \vert^{-1}$, and it does not depend on $j$.
In contrast, $\langle r \rangle^{(3)}_{j}$ depends on $j$ when the contribution
of the dyon electric charge is taken into account.
The reason is that the  dyon's  electric  charge  (\ref{III:2}) is proportional
to $q$, and the  angular  momentum $j = \left\vert q \right\vert - 1/2$ for the
fermionic states of the third type.
In Fig.~\ref{fig4}, the parameter  $q  =  1$,  which corresponds to the angular
momentum $j = 1/2$.
In this case, the $z$ projection $m= -1/2,1/2$ and the electric dipole momentum
is different from zero.

From Fig.~\ref{fig4}  it follows that  taking  into account the dyon's electric
charge leads to a  decrease  in  the  mean radius of the fermion  distribution,
which results in  a  decrease  in  the magnitude of the electric dipole moment.
The reason is that the electric potential  (\ref{III:2a}) is attractive for all
values of $\theta$ in Fig.~\ref{fig4}.
Further, it follows  from  Fig.~\ref{fig4}  that the correction  to  the dipole
moment  (\ref{V:4})  caused    by    the   electric   charge   of  the  dyon is
insignificant when the state (\ref{II:36}) is tightly bound.
In contrast, these corrections  become  substantial  for  loosely bound states,
i.e. for  $\left\vert \theta \right\vert  = \pi/2 + \epsilon$,  where $\epsilon
\ll 1$.
This is because in this region of  $\theta$,  the  contribution of the electric
potential  $A_{0}$  to  the  forming  of  the  fermionic  bound  state  becomes
comparable to the contribution of the boundary condition (\ref{II:33}).

It was shown in subsection  \ref{subsec:IVA}  that  in  addition  to  the state
(\ref{II:36}), which is tightly bound in  general, there  is  also  an infinite
sequence of loosely bound states.
Being fermionic  states  of  the  third type, they possess the minimum possible
angular momentum $j = \left\vert q \right \vert - 1/2$.
To find the mean radii $\left\langle r\right\rangle^{(3)}_{nj}$ of these states,
we use the numerical methods of subsection \ref{subsec:IVA}.
First, we find the combination $F_{nj}\left(r\right) = f_{nj}\left(r\right) - i
\tan\left[\theta/2 + \pi/4\right] g_{nj} \left( r \right)$  that  satisfies the
homogeneous Dirichlet boundary conditions.
Knowing $F_{nj}$ and using the system (\ref{II:23}), we  can express the radial
wave functions $f_{nj}$ and $g_{nj}$ as linear combinations of $F_{nj}$ and its
derivative $F'_{nj}$, and then  calculate  the corresponding mean radius $\left
\langle r\right\rangle^{(3)}_{nj}$.

\begin{figure}[tbp]
\includegraphics[width=0.5\textwidth]{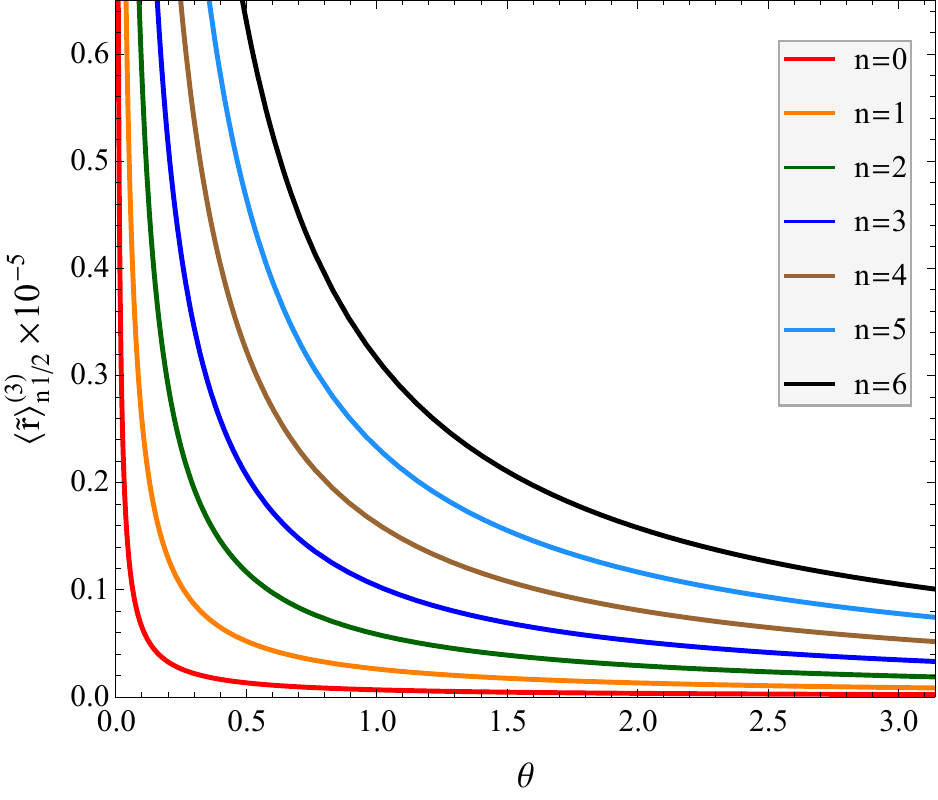}
\caption{\label{fig5}    Dependence  of  the dimensionless  mean  radius $\left
\langle \tilde{r}\right\rangle^{(3)}_{n 1/2}= M\left\langle r\right\rangle^{(3)
}_{n 1/2}$ on $\theta$ for the  first  few  values of the radial quantum number
$n$. The curves correspond to the parameters $\alpha = 1/137$ and $q = 1$}
\end{figure}

Figure~\ref{fig5} presents  the  dependence  of  the dimensionless  mean radius
$\left\langle \tilde{r} \right\rangle^{(3)}_{n 1/2}  =  M \left\langle r \right
\rangle^{(3)}_{n 1/2}$ on the parameter $\theta$ for the  first  few  values of
the radial quantum number $n$.
It was found that in Fig.~\ref{fig5}, the curves $\left\langle \tilde{r} \right
\rangle^{(3)}_{n 1/2}$ can be approximated  by the hyperbolas
\begin{equation}
\left\langle \tilde{r}\right\rangle_{n 1/2}^{\left(3\right)}
\approx a_{n 1/2}^{\left(3\right) }\theta ^{-1}                    \label{V:4b}
\end{equation}
with an accuracy of $\sim 0.1 \%$.
The  coefficient    $a_{n 1/2}^{\left( 3\right)}$    increases    monotonically
(approximately $\propto n(n+b)$, where $b$ is a constant) with  an  increase in
the radial quantum number $n$.
Note that the radial quantum number  $n$ is equal to the number of zeros of the
function $F_{nj}= f_{nj} - i\tan\left[\theta/2+ \pi/4\right] g_{nj}$ except for
the zero at $r = 0$.
At the same time, all zeros of the radial  wave functions $f_{nj}$ and $g_{nj}$
are at nonzero $r$, and there are $n+1$ of them.
The behaviour and  the  scale  of  the curves in Fig.~\ref{fig5} are similar to
those for  the  curves  shown in Fig.~\ref{fig6},  and will be explained in the
next subsection.

\subsection{\label{subsec:VB}          Electric  dipole  moments  of  the bound
fermionic states  with  the angular momentum $j \ge \left\vert q \right \vert +
1/2$}

It was shown in \cite{kzm3_prd_1977}  that  in  this  case, the electric dipole
moment is
\begin{equation}
\left\langle d_{z}\right\rangle _{n j m}=-\frac{mq}{j\left( j+1\right) }
e\left\langle r\right\rangle _{n j},                                \label{V:5}
\end{equation}
where the mean radius of the fermion distribution
\begin{equation}
\left\langle r\right\rangle _{n j}=\int\nolimits_{0}^{\infty }r\bigl(
\left\vert f_{n j}\right\vert ^{2}+\left\vert g_{n j}\right\vert
^{2}\bigr) dr                                                       \label{V:6}
\end{equation}
and the $z$ projection $m\in \left[ -j,j\right]$.
In  contrast  to  the  previous  case $j = \left\vert q \right\vert - 1/2$, the
dipole moment (\ref{V:5}) differs  from  zero  even if $q = \pm 1/2$, since the
angular momentum $j$ is now greater than zero. 
Eq.~(\ref{V:5}) tells  us  that  all  dependence  of  $\left\langle d_{z}\right
\rangle _{njm}$ on $\theta$  is  contained  in  the mean radius $\left\langle r
\right\rangle_{nj}$.

\begin{figure}[tbp]
\includegraphics[width=0.5\textwidth]{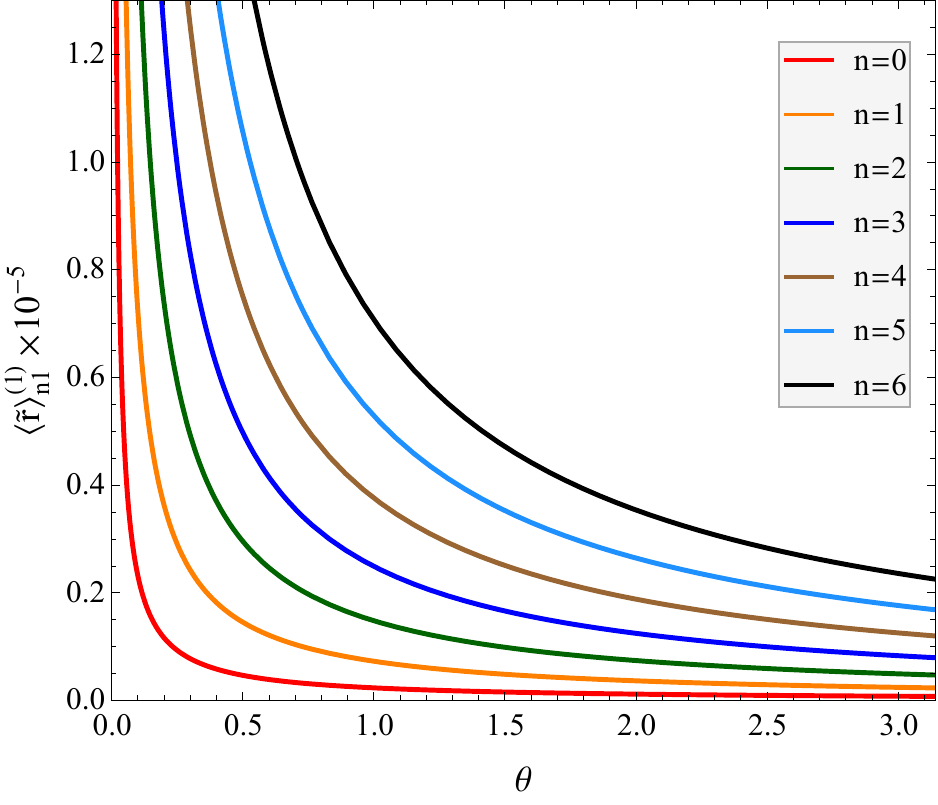}
\caption{\label{fig6}    Dependence  of  the dimensionless  mean  radius $\left
\langle\tilde{r}\right\rangle^{(1)}_{nj}=M\left\langle r\right\rangle^{(1)}_{nj
}$ on $\theta$ for the angular momentum $j=1$ and the radial quantum numbers $n
= 0, \ldots, 6$.   The curves correspond to the parameters $\alpha = 1/137$ and
$q = 1/2$}
\end{figure}

Figure~\ref{fig6} presents  the  dependence  of  the  dimensionless mean radius
$\left\langle\tilde{r}\right\rangle^{(1)}_{n 1}=M\left\langle r\right\rangle^{(
1)}_{n 1}$ on $\theta$ for the radial quantum numbers $n = 0, \ldots, 6$.
The results  correspond  to  the   bound   fermionic  states  of the first type
$\psi^{(1)}_{n1m}$.
All the curves in Fig.~\ref{fig6} are approximated  with  high  accuracy by the
hyperbolas  $a^{(1)}_{n1}/\theta$,   where   the   coefficients  $a^{(1)}_{n1}$
increase monotonically with an increase in $n$.
The same pattern also holds for fixed $n$ and increasing $j$.
In  this   case,   the    curves    are    approximated   by   the   hyperbolas
$a^{(1)}_{nj}/\theta$   with  the  coefficients  $a^{(1)}_{n j}$  monotonically
increasing with an increase in $j$.
Similar results  hold  for  the  bound  fermionic  states  of  the  second type
$\psi^{(2)}_{njm}$.
In particular, in this case, the mean radius $\left\langle r \right\rangle^{(2)
}_{nj}$ is also inversely proportional  to the parameter $\theta$ and increases
monotonically with an increase in $n$ or $j$.

The  hyperbolic  $\theta$-dependence  of  the  curves  in  Fig.~\ref{fig6}  can
be obtained from the approximate  solution (\ref{IV:15}) -- (\ref{IV:20}).
Indeed, by substituting $r = \varkappa_{n j}^{-1} \rho$ in integral (\ref{V:6})
and using  Eqs.~(\ref{IV:15}) -- (\ref{IV:20}),   we   can   show  that  $\left
\langle   r  \right  \rangle^{(1)}_{n j}  \propto  \varkappa_{n j}^{-1}$  up to
corrections of order $\alpha^{2}$.
We see that in the leading order in $\alpha$, the dependence of $\left\langle r
\right\rangle^{(1)}_{nj}$   on   the    parameter  $\theta$   is   isolated  in
$\varkappa_{nj}$.
Using Eq.~(\ref{IV:17}) and the definition $\varkappa_{nj}=\left( M^{2} - E_{nj
}^{2}\right)^{1/2}$,  we  find   that  $\varkappa_{nj}^{-1} = \pi \left(n + \mu
\right)/\left( M \alpha q \theta \right) +O\left[\alpha \right]$.
Hence, in the leading order in $\alpha$, $\varkappa_{nj}^{-1} \propto\theta^{-1
}$, which  explains   the   hyperbolic  $\theta$-dependence  of  the  curves in
Fig.~\ref{fig6}.

Using the approximate solution (\ref{IV:15}) -- (\ref{IV:20}),  one  can obtain
the asymptotics of $\left\langle r \right\rangle_{nj}$ for large $n$ or $j$.
In particular, for large $n$ and fixed $j$, we find  that
\begin{equation}
\left\langle r\right\rangle _{n j}^{\left( a\right) }\sim
\bigl(b n + c^{\left(a\right)}\left( \mu \right)\bigr)
\left( n+\mu \right)/\left( Mz\alpha \right),                       \label{V:7}
\end{equation}
where $b$ is a positive constant  of  the  order  of unity, $c^{\left(a\right)}
\left( \mu \right)$  is  an  increasing  quasilinear  function of the parameter
$\mu = \bigl[\left(j+1/2\right)^{2}-q^{2}\bigr]^{1/2}$, $z = q \theta/\pi$, and
the index $a = 1\,(2)$ for the fermionic states of the first (second) type.
Similarly, for large $j$ and fixed $n$, we find that
\begin{equation}
\left\langle r\right\rangle _{nj}^{\left( a\right) } \sim
j \left(j+3n+a+1/2\right) /\left( Mz\alpha \right).                  \label{V:8}
\end{equation}

A comparison of Figs.~\ref{fig4} and  \ref{fig6}  shows  that  the  mean radius
$\langle r\rangle^{(1)}_{n j}$  significantly (approximately by four  orders of
magnitude) exceed the mean  radius $\langle r\rangle^{(3)}$  of the bound state
(\ref{II:36}).
The reason is that the existence of the bound state  (\ref{II:36}) is caused by
the boundary  condition  (\ref{II:33}), whereas  the  existence   of  the bound
states $\psi^{(1)}_{njm}$ is entirely due to the electric charge of the dyon.
As a result, the mean  radius  $\left\langle r \right\rangle^{(1)}_{nj} \approx
F^{(1)}\left(n,j\right)\varkappa_{nj}^{-1}\approx F^{(1)}\left(n,j\right)\times$
$\left(n+\mu\right)\pi/\left(M \alpha q \theta\right)$, where $F^{(1)}\left(n,j
\right)$ is a function of $n$ and $j$.
We see that $\left\langle r \right\rangle^{(1)}_{n j}$  is  proportional to the
large factor  $\pi/(\alpha q)\approx  860$,  whose  large  magnitude  is due to
the smallness of the fine structure constant $\alpha = 1/137$.
The remaining factor  $F^{(1)}\left(n, j\right)\left(n + \mu\right)$  is on the
order of $10 \div 20$.
In contrast, for the unperturbed  bound  state (\ref{II:36}),  the  mean radius
$\left\langle  r  \right\rangle^{(3)}   =  \left(2 M \left\vert\cos\left(\theta
\right)\right\vert\right)^{-1}$, and accounting for the Coulomb correction does
not lead to a qualitative change in the situation.
We see that in contrast to $\left\langle r \right\rangle^{(1)}_{n j}$, the mean
radius $\left\langle r\right\rangle^{(3)}$ does not contain a  large multiplier
$\propto \alpha^{-1}$.
This explains  the  difference  in  the  magnitudes  of  $\left\langle r \right
\rangle^{(1)}_{n j}$ and $\left\langle r\right\rangle^{(3)}$.
The difference in the magnitudes of $\left\langle r \right \rangle^{(2)}_{n j}$
and $\left\langle r\right\rangle^{(3)}$ is explained similarly.

Due to  the  nonzero  electric  dipole   moments,  the   energy  levels of  the
fermion-dyon  system  are  split  in  an  external  homogeneous  electric field
(Stark effect).
This splitting removes the degeneracy with respect to the $z$-projection $m$ of
the angular momentum.
Because of the difference in the mean radii $\left\langle r \right \rangle^{(1,
2)}_{n j}$  and  $\left\langle  r \right\rangle^{(3)}$,  the  splitting  of the
tightly bound levels, which is possible only if $\left\vert q \right\vert>1/2$,
will be much smaller than that of the loosely bound levels.

\section{Magnetic dipole moments of the bound fermionic states}  \label{sec:VI}

The operator of magnetic moment is
\begin{equation}
\boldsymbol{\mu}=\frac{e}{2}\mathbf{r}\!\times\!\boldsymbol{\alpha}.
                                                                   \label{VI:1}
\end{equation}
It follows from Eq.~(\ref{VI:1})  that  $\boldsymbol{\mu}$  is  an axial-vector
operator.
Using Eqs.~(\ref{II:20}) and (\ref{VI:1}), it is easy to show that the magnetic
dipole moment  of  the   fermionic   states   of   the   third   type  vanishes
\cite{kzm3_prd_1977}.
Indeed, there  exists only  one two-component angular eigensection $\eta_{q m}$
when $j = \left\vert q \right \vert - 1/2$.
As a result, the two cross terms produced  by  the matrix $\boldsymbol{\alpha}$
in $\boldsymbol{\mu}$, have  the  same  angular structure $\eta_{q m}^{\dagger}
\left( \mathbf{r}\!\times\!\boldsymbol{\sigma }\right) \eta_{qm}$.
It is known \cite{kzm1_prd_1977} that $\eta_{q m}$  is  an  eigensection of the
operator $\mathbf{r}\! \cdot\! \boldsymbol{\sigma}$    with    the   eigenvalue
$q/\left\vert q \right\vert$.
At the  same  time,  the  operator  $\mathbf{r} \!\cdot \! \boldsymbol{\sigma}$
anticommutes with the operator $\mathbf{r}\! \times \!\boldsymbol{\sigma}$.
It follows that the dipole magnetic  moment $\boldsymbol{\mu}$ vanishes for the
bound fermionic states of the third type.

In contrast, the magnetic dipole  moment  is  different from zero for the bound
states of the first and second types.
It was shown  in  \cite{kzm3_prd_1977} that in this case, the  magnetic  dipole
moment is
\begin{equation}
\left\langle \mu _{z}\right\rangle _{njm}^{\left( a\right) }=\left(
-1\right) ^{a+1}\frac{e }{2}\frac{m \mu}{j\left( j+1\right) }
\int\nolimits_{0}^{\infty }r\left( if^{\ast}_{nj} g_{nj}\right) dr,\label{VI:2}
\end{equation}
where $\mu= \bigl[ \left(j + 1/2\right)^{2} - q^{2}\bigr]^{1/2}$, and the index
$a = 1\, (2)$ for a state of the first (second) type.
Numerical calculations reveal that the dipole moment $\left\langle\mu_{z}\right
\rangle_{njm}^{\left( a\right) }$  is  practically independent of the parameter
$\theta$.
Analytical  calculations  performed  with the use of  approximate  solutions of
Section~\ref{sec:IV} show that in the  leading  order  in $z\alpha$, the dipole
moment
\begin{equation}
\left\langle \mu _{z}\right\rangle _{njm}^{\left( a\right) }\approx
\frac{e}{8 M}\frac{m\mu }{j\left( j+1\right) }\bigl( 1+\left( -1\right)
^{a+1}2\mu \bigr).                                                 \label{VI:3}
\end{equation}
Eq.~(\ref{VI:3})  is  independent  of  $\theta$,  and  describes  the numerical
results with high accuracy.

The  absence   of   the   $\theta$-dependence  of  the  dipole  magnetic moment
$\left\langle\mu_{z} \right\rangle_{njm}^{\left(a\right)}$  is explained by the
presence of the cross term $f^{*}g$ in Eq.~(\ref{VI:2}).
Indeed, the system  of  equations (\ref{IV:6a}) and (\ref{IV:6b}) tells us that
in  leading  order  in  $\alpha$, the radial  function $g_{nj} \approx i \left(
f^{\prime}_{nj}\mp\mu f_{nj}/r\right)/\left(2M\right)$, where the upper (lower)
sign corresponds to a state of the first (second) type.
Further, Eqs.~(\ref{IV:15}) and  (\ref{IV:17})  tell  us  that  the approximate
solution $f_{n j}$ depends on $r$  only  through  the dimensionless combination
$\rho = \varkappa_{n j} r$.
Expressing the radial function  $g_{n j}$  in  terms of $f_{nj}$ and making the
substitution $r=\varkappa_{nj}^{-1}\rho$, we conclude that in Eq.~(\ref{VI:2}),
the dependence of the integral $\int \nolimits_{0}^{\infty }r \left(i f^{\ast}g
\right) dr$        on       $\theta$     is    isolated      in     the  factor $\mathcal{N}_{nj}^{2}\varkappa_{nj}^{-1}$.
Using the formulae of Section~\ref{sec:IV}, it is easy to show that this factor
does not depend on either $\theta$ or $\alpha$.
This explains the independence of the dipole moment $\left\langle \mu_{z}\right
\rangle_{njm}^{\left( a \right)}$ of $\theta$  in  the  leading (zero) order in
$\alpha$.

It follows from the above that  the  $\theta$-dependences of the dipole moments
$\left\langle d_{z}\right\rangle_{njm}^{\left(a\right)}\propto \theta^{-1}$ and
$\left\langle\mu_{z}\right\rangle_{n j m}^{\left( a\right)} \propto \theta^{0}$
are completely different.
Furthermore, numerical calculations  and  Eq.~(\ref{VI:3}) tell  us that $\left
\langle\mu_{z} \right  \rangle_{n j m}^{\left( a \right)}$   is  independent of
the radial quantum number $n$.
In  contrast,  it   follows   from   Fig.~\ref{fig6}  and  Eqs.~(\ref{V:7}) and
(\ref{V:8})  that  $\left \langle  d_{z} \right \rangle_{njm}^{\left(a\right)}$
increases monotonically with an increase in $n$.
At the same time, Eqs.~(\ref{V:5}), (\ref{V:8}),  and (\ref{VI:3}) tell us that
for  large  values  of  $j$,   both   $\left\langle\mu_{z}\right\rangle_{njm}^{
\left(a\right)}$ and $\left\langle d_{z} \right \rangle_{njm}^{\left(a\right)}$
cease to depend on $j$ while remaining proportional to the $z$ projection $m$.
Finally, in the leading order in $\alpha$,  the  electric  dipole moment $\left
\langle d_{z} \right \rangle_{n j m}^{\left( a \right)}  \propto  \alpha^{-1}$,
whereas the  magnetic  dipole  moment $\left\langle\mu_{z}\right\rangle_{njm}^{
\left(a\right)} \propto \alpha^{0}$.
As a result, for  a  given  state, the magnitude of the  electric dipole moment
significantly exceeds the magnitude of the magnetic dipole moment.
These significant differences between  $\left\langle d_{z}\right\rangle_{njm}^{
\left(a\right)}$ and $\left\langle\mu_{z}\right \rangle_{njm}^{\left(a\right)}$
are due to the different structures of the radial integrals in Eqs.~(\ref{V:5})
and (\ref{VI:2}).

Like a homogeneous electric field, a  homogeneous  magnetic  field  removes the
degeneracy of the energy  levels of the fermion-dyon system with respect to the
quantum number $m$ (Zeeman effect).
The splitting  is only  possible  for  the  states  with  the  angular  momenta
$j\ge \left\vert q\right\vert+1/2$, since the magnetic dipole moment of a bound
fermionic state vanishes otherwise.

\section{Conclusion}                                            \label{sec:VII}

In this paper, we have studied the bound fermionic states in the external field
of an Abelian dyon, which is the  Dirac  monopole surrounded by the cloud of an
induced electric charge.
The  fermion-monopole   system     has    several   characteristic  properties.
In particular, the magnetic  field  of   the  monopole  changes  sign under $P$
transformation, which is equivalent  to  the  change  of sign of the monopole's
magnetic charge.
Since $C$ conjugation also  changes the sign of the monopole's magnetic charge,
combined $CP$ transformation leaves the  monopole's   magnetic field unchanged.
From this,  one  could  naively  conclude   that  $CP$  is  a  symmetry  of the
fermion-monopole system.
However, there is one obstacle that makes this impossible.
The absence of the centrifugal barrier  in  the states with the minimal angular
momentum $j = \left\vert q \right\vert - 1/2$  leads to the fact that the Dirac
Hamiltonian is not self-adjoint on these states.

To solve this problem, a boundary condition must be imposed so  that  the Dirac
Hamiltonian possesses a complete set of  eigenfunctions.
The appropriate boundary  condition  at $r = 0$ forms a one-parameter family of
self-adjoint  extensions  of  the  Dirac  Hamiltonian \cite{goldhaber_prd_1977,
 callias_prd_1977}.
Thus,  in  addition  to  the   Hamiltonian,   the   fermion-soliton  system  is
characterized  by  an  angular   parameter  $\theta$ entering into the boundary
condition.
The parameter $\theta$ defined  modulo $2 \pi$  results in the existence of the
$\theta$ vacua.

Although $CP$  leaves the magnetic  field of the monopole unchanged, it changes
sign of the parameter $\theta$.
In the  case  of  massive  fermions,   this  results  in  a  breaking  of  $CP$
invariance of the fermion-monopole system. 
This makes possible  the  appearance  of  the  electric  charge of the monopole
via  quantum  effects  (polarization  of   the  fermionic  vacuum),  since  the
spherically symmetric electric  field  is  also not invariant (it changes sign)
under $CP$.

Indeed, it has been shown in \cite{yamagishi_prd_1983} that the polarization of
the fermionic vacuum in the vicinity   of   the  Dirac  monopole  leads  to the
appearance  of  a  spherically  symmetric  electric charge density.
Because of  the  mild  integrable  singularity  $\propto r^{-2}\ln(M r)$ of the
electric charge density at the origin, the electrostatic energy of the electric
charge distribution is finite.
The electric   charge   for   a   unit   monopole   obeys   the  Witten formula
$Q = -e \theta/2\pi$ so that the monopole becomes a dyon.
At spatial infinity, the dyon's electric potential is  asymptotically Coulombic
($\sim Q/r$),  but  it  differs  significantly  from  the  Coulomb potential when
$r \lesssim M^{-1}$.

The nontrivial  boundary  condition  at  $r = 0$  results  in a bound fermionic
state \cite{goldhaber_prd_1977, callias_prd_1977}    possessing   the   minimal
angular momentum $j = \left\vert q \right\vert - 1/2$.
In the general case, this state is tightly bound.
The consideration of the  dyon's  electric field leads to a shift in the energy
of this bound state  resulting  in  an increase in the magnitude of the binding
energy.

In addition to perturbing the already existing tightly bound  state, the dyon's
electric field leads to the appearance of new bound states.
These states are loosely bound, and  there  are  an infinite number of them for
each value  of  the  angular  momentum, including  the  minimum  possible value
$j = \left\vert q \right\vert - 1/2$.
In the latter case, to  ensure  the  Hermiticity  of the Dirac Hamiltonian, the
fermionic  wave  functions  must  satisfy the boundary condition (\ref{II:33}),
which remains  unchanged  due  to  the  the weak logarithmic singularity of the
electric potential.
At the same  time,  the  point  distribution  of  the  electric  charge density
$\rho(\mathbf{r})=Q\delta(\mathbf{r})$ leads to a falling of the fermion on the
centre, which is unacceptable from a physical viewpoint.
Furthermore, the boundary condition (\ref{II:33}) is  incompatible  with such a
behaviour of the fermion.

Because of the centrifugal barrier, there is  no  need for a boundary condition
at $r = 0$ for the states with the nonminimal angular momenta $j \ge \left\vert
q \right \vert + 1/2$.
In this case, an analytical approximation  of the  potential is possible, which
allows us to find the solution to  the  Dirac  equation  in an analytical form.
The approximated analytical spectrum  is  hydrogen-like and resembles the exact
analytical   spectrum    for    a    purely    Coulomb    fermion-dyon   system
\cite{zhang_prd_1986,zhang_prd_1989,zhang_jmp_1990,zhang_prd_1990}.

Since the  fermion-dyon  system  is  not  invariant  under  $P$ transformation,
all its bound states  have nonzero electric dipole moments, which distinguishes
it from the hydrogen atom.
These  electric  dipole  moments depend nontrivially on the parameter $\theta$.
The  analytical  approximate   solution   for   the   bound   states   with the
angular  momenta  $j \ge \left\vert q \right\vert  +  1/2$ makes it possible to
find the asymptotics of  the electric  dipole  moments for large values of $j$.
In the most interesting  case  of $\left\vert q \right\vert = 1/2$, the minimum
angular momentum $j = \left\vert q \right\vert- 1/2$ vanishes, which results in
the vanishing of the dipole moment of the corresponding bound state.

The bound  states  with  nonminimal  angular  momenta  possess nonzero magnetic
dipole moments.
In contrast, the magnetic dipole moments  vanish  for the bound states with the
minimal angular momentum $j = \left\vert q \right\vert- 1/2$, which may ($\left
\vert q \right\vert=1/2$) or may not ($\left\vert q \right\vert >1/2$) be zero.
Unlike the electric dipole moments, the magnetic dipole moments are practically
independent of the parameter $\theta$.
Also, for a given state,  the  magnitude  of the electric dipole moment is much
larger than that of the magnetic dipole moment.
These  differences  are  caused  by  the  different  structures  of  the  radial
integrals determining the values of these two moments.

\begin{acknowledgements}

This work was supported by the Russian Science Foundation, grant No 23-11-00002.

\end{acknowledgements}

\end{document}